\documentclass{aa}  
\usepackage{graphicx}
\usepackage{txfonts}
\usepackage[]{hyperref}
\hypersetup{colorlinks=false}
\usepackage[utf8]{inputenc}
\usepackage[english]{babel}
\usepackage{url}
\usepackage{amssymb}
\usepackage{color}
\usepackage{multirow}
\usepackage{todonotes}
\usepackage{soul}
\usepackage{ulem}

\newcommand{\emm}[1]{\ensuremath{#1}}
\newcommand{\emr}[1]{\emm{\mathrm{#1}}}
\newcommand{\chem}[1]{\emr{\,#1}} 

\newcommand{\unit}[1]{\emr{\,#1}}
\newcommand{\kms}{\unit{km\,s^{-1}}}

\newcommand{\Jone}{(1--0)}
\newcommand{\Jtwo}{(2--1)}
\newcommand{\Ht}{\emr{H_2}}

\newcommand{\HCOp}{\chem{HCO^{+}}}
\newcommand{\HCSp}{\chem{HCS^{+}}}

\newcommand{\HHCO}{\chem{H_{2}CO}}

\newcommand{\CCH}{\chem{C_{2}H}}
\newcommand{\lCCCH}{\chem{l-C_{3}H}}
\newcommand{\cCCCH}{\chem{c-C_{3}H}}
\newcommand{\CCCHp}{\chem{C_{3}H^{+}}}
\newcommand{\lCCCHH}{\chem{l-C_{3}H_{2}}}
\newcommand{\cCCCHH}{\chem{c-C_{3}H_{2}}}
\newcommand{\metcy}{\chem{CH_{3}CN}}

\newcommand{\CHHH}{\chem{CH_{3}}}

\newcommand{\rev}[1]{{\bf{#1}}}

  \newcommand{\TabLine}{
  \begin{table*}
    \centering %
    \caption{Detected lines.}
    \begin{tabular}{lcccccr}
  \hline \hline
 Molecule$^a$ & Transition & Frequency & A & E$_u$     & N/$\int \tau dv$$^{b}$   & rms$^{c}$ \\
         &         &  (GHz)              & (10$^{-5}$ s$^{-1}$)     & (K) & (10$^{12}$cm$^{-2}$km$^{-1}$s) & ($10^{-2}$ )   \\
  \hline
 o-H$_2$CO &$2_{1,1} - 1_{1,0} $         & 150.4983340 & 6.47 & 7.46 & 7.51            &  0.37 \\
 p-H$_2$CO  & $2_{0,2} - 1_{0,1} $       &145.6029490 & 7.81 & 10.48 & 5.58        & 0.38 \\
 o-H$_2$$^{13}$CO & $2_{1,1} - 1_{1,0} $ & 146.6356717  & 5.99 & 7.26 & 7.54 & 0.26 \\
 CS &         $3 - 2$                                  & 146.9690287 & 6.07 & 14.11 & 26.9      & 0.28\\
 C$^{34}$S & $3 - 2$ &                              144.6171007 & 5.74 & 13.88 & 26.4     & 0.29 \\
 o-c-C$_3$H$_2$ & $3_{1,2} - 2_{2,1}$ & 145.0896055 & 6.77 & 13.71 & 54.0     & 0.36 \\
 o-c-C$_3$H$_2$ & $4_{1,4} - 3_{0,3}$ & 150.8519080 &16.37 & 16.97 & 57.7      &0.34 \\  
 p-c-C$_3$H$_2$ & $2_{2,0} - 1_{1,1}$ & 150.4365547 & 5.36 & 9.71 & 9.99    & 0.37\\
  \hline
  \end{tabular} 
   \tablefoot{
   \tablefoottext{a}{Spectroscopic data have been taken from the Cologne Data Base for Molecular Spectroscopy (CDMS) \citep{Muller:2005,Endres:2016}}
\tablefoottext   {b}{Assuming an excitation temperature of 2.73~K}
\tablefoottext {c}{rms on the line to continuum determined on the normalized spectra}
   }
    \label{tab:lines}
  \end{table*}
  }

\newcommand{\Tabfit}{%
  \begin{table*}
    \centering %
    \caption{Properties of the velocity components.}
\begin{tabular}{lcccccc}
\hline \hline
Line & Velocity & FWHM & $\int \tau dv$ & $\tau_{max}$ & T$_{ex}$ & N$^a$ \\
         &       (\kms) & (\kms) &               (\kms)     &                    & (K)        & (10$^{12}$ cm$^{-2}$ ) \\
\hline
CS ($2-1$) & $-17.2$ & $0.41\pm 0.02$ &     $0.42 \pm 0.02$ & $0.96 \pm 0.04$ & $\sim 3.0$ & $3.3 \pm 0.3 $ \\
CS ($2-1$) & $-13.8$ &                          &     $<0. 1$             & $<0.15$              &  2.73 &  $<0.33$ \\
 CS ($2-1$) & $-10.3$ & $0.39\pm 0.01$ &     $0.76 \pm 0.03$ & $1.8 \pm 0.04$ & $\sim 3.3$   & $6.4 \pm 0.3 $ \\
 CS ($2-1$) & $-8.5$ & $2.18\pm 0.8$ &     $0.22\pm 0.06$ & $0.10 \pm 0.04$ & 2.73 & $1.3 \pm 0.4 $ \\
 CS ($2-1$) & $-4.1$ & $1.8\pm 0.5$ &     $0.124\pm 0.04$ & $0.06 \pm 0.04$ & 2.73 & $0.5 \pm 0.3 $ \\  
\hline
CS ($3-2$) & $-17.2$ & $0.37\pm 0.01$ &     $0.15 \pm 0.01$ & $0.38 \pm 0.01$ &   $\sim 3.0$          & $3.3 \pm 0.3 $ \\
CS ($3-2$) & $-13.8$ &                          &     $<0.005 $           & $<0.011$               & 2.73  &  $<0.33$ \\
 CS ($3-2$) & $-10.3$ & $0.49\pm 0.01$ &     $0.41 \pm 0.01$ & $0.78 \pm 0.01$ &   $\sim 3.1$           & $6.4 \pm 0.3 $ \\
 CS ($3-2$) & $-8.5$ & $1.4\pm 0.10$ &     $0.036\pm 0.005$ & $0.024 \pm 0.01$ & 2.73 & $1.3 \pm 0.4 $ \\
 CS ($3-2$) & $-4.1$ & $1.5\pm 0.17$ &     $0.019\pm 0.003$ & $0.011 \pm 0.01$ & 2.73 & $0.5 \pm 0.3 $ \\
 \hline
  C$^{34}$S ($3 - 2$) & $-17.2$ & $0.33 \pm 0.09$ & $0.0056 \pm 0.001$ & $0.016 \pm 0.003$ &  $3.0$ & $0.13 \pm 0.03 $\\
    C$^{34}$S ($3 - 2$) & $-10.4$ & $0.59 \pm 0.05$ & $0.016 \pm 0.001$ & $0.026 \pm 0.003$ & $3.3$ & $0.34 \pm 0.03$\\
 \hline
 p-H$_2$CO ($2_{0,2} - 1_{0,1} $) & $-17.2$ & $0.54 \pm 0.01$ & $0.11 \pm 0.01$ & $0.19 \pm 0.007 $ & $2.73$ & $0.62 \pm 0.06 $ \\
 p-H$_2$CO ($2_{0,2} - 1_{0,1} $) & $-13.8$ & $1.7 \pm 0.1$ & $0.05\pm 0.01$ & $0.03 \pm 0.007 $ & $2.73$ & $0.28 \pm 0.05 $ \\
 p-H$_2$CO ($2_{0,2} - 1_{0,1} $) & $-10.4$ & $0.43 \pm 0.01$ & $0.19 \pm 0.01$ & $0.42 \pm 0.007 $ & $2.73$ & $1.09 \pm 0.06 $ \\
 p-H$_2$CO ($2_{0,2} - 1_{0,1} $) & $-8.4$ & $0.97\pm 0.04$ & $0.09 \pm 0.01$ & $0.08\pm 0.007 $ & $2.73$ & $0.48 \pm 0.06 $ \\
 p-H$_2$CO ($2_{0,2} - 1_{0,1} $) & $-4.2$ & $1.5\pm 0.2$ & $0.05\pm 0.01$ & $0.03 \pm 0.007 $ & $2.73$ & $0.28 \pm 0.05 $ \\
 \hline
 o-H$_2$CO ($2_{1,1} - 1_{1,0} $) & $-17.2$ & $0.53 \pm 0.01$ & $0.27 \pm 0.01$ & $0.47\pm 0.01 $ & $2.73$ & $2.2 \pm 0.1 $ \\
 o-H$_2$CO ($2_{1,1} - 1_{1,0} $)  & $-13.8$ & $1.6 \pm 0.1$ & $0.11\pm 0.01$ & $0.07 \pm 0.01 $ & $2.73$ & $0.84 \pm 0.05 $ \\
 o-H$_2$CO ($2_{1,1} - 1_{1,0} $) & $-10.4$ & $0.44 \pm 0.01$ & $0.42 \pm 0.01$ & $0.89 \pm 0.01$ & $2.73$ & $3.5 \pm 0.2 $ \\
 o-H$_2$CO ($2_{1,1} - 1_{1,0} $)  & $-8.4$ & $0.97\pm 0.02$ & $0.19 \pm 0.01$ & $0.018\pm 0.01$ & $2.73$ & $1.4\pm 0.1 $ \\
 o-H$_2$CO ($2_{1,1} - 1_{1,0} $)  & $-4.2$ & $1.6\pm 0.06$ & $0.11\pm 0.01$ & $0.06 \pm 0.01$ & $2.73$ & $0.84 \pm 0.05 $ \\
 \hline
  o-H$_2$$^{13}$CO ( $2_{1,1} - 1_{1,0} $ ) &  $-10.4$ &  $0.71 \pm 0.1$ &  $0.0076 \pm 0.001$ & $0.01 \pm 0.002$ & $2.73$ & $0.057 \pm 0.01$ \\
 \hline
   o-c-C$_3$H$_2$ ( $3_{1,2} - 2_{2,1}$) & $-17.2$ & $0.41 \pm 0.1 $ & $ 0.0063 \pm 0.001$ & $0.014 \pm 0.004$ & 2.73 & $0.34 \pm 0.07$   \\
   o-c-C$_3$H$_2$ ( $3_{1,2} - 2_{2,1}$) & $-13.6$ & $1.34 \pm 0.4 $ & $ 0.0088 \pm 0.002$ & $0.006 \pm 0.004$ & 2.73 & $0.47 \pm 0.1$   \\
   o-c-C$_3$H$_2$ ( $3_{1,2} - 2_{2,1}$) & $-10.4$ & $0.57  \pm 0.1 $ & $ 0.019 \pm 0.002$ & $0.03 \pm 0.004$ & 2.73 & $1.0 \pm 0.1$   \\
   o-c-C$_3$H$_2$ ( $3_{1,2} - 2_{2,1}$) & $-8.6$ & $1.32 \pm 0.2 $ & $ 0.021 \pm 0.003$ & $0.015 \pm 0.004$ & 2.73 & $1.1 \pm 0.16$   \\
    o-c-C$_3$H$_2$ ( $3_{1,2} - 2_{2,1}$) & $-4.0$ & $1.81 \pm 0.4 $ & $ 0.012 \pm 0.003$ & $0.006 \pm 0.004$ & 2.73 & $0.65 \pm 0.16$   \\
   \hline
   o-c-C$_3$H$_2$ ($4_{1,4}-3_{0,3}$) &  $-17.2$ & $0.5\pm 0.1$ & $0.0068 \pm 0.001$ & $0.013 \pm 0.003$ & 2.73 & $0.39 \pm 0.09$ \\  
   o-c-C$_3$H$_2$ ($4_{1,4}-3_{0,3}$) &  $-10.4$ & $0.62\pm 0.1$ & $0.016 \pm 0.002$ & $0.024 \pm 0.003$ & 2.73 & $0.92 \pm 0.1$ \\  
   o-c-C$_3$H$_2$ ($4_{1,4}-3_{0,3}$) &  $-8.5$ & $1.7\pm 0.2$ & $0.024 \pm 0.003$ & $0.013 \pm 0.003$ & 2.73 & $1.4 \pm 0.2$ \\  
   o-c-C$_3$H$_2$ ($4_{1,4}-3_{0,3}$) &  $-3.8$ & $2.0\pm 0.3$ & $0.014 \pm 0.002$ & $0.006 \pm 0.003$ & 2.73 & $0.81 \pm 0.12$ \\  
    \hline
 p-c-C$_3$H$_2$ ($2_{2,0}-1_{1,1}$) & $-17.2$ & $0.73 \pm 0.13$ & $0.016 \pm 0.002$ & $0.021 \pm 0.004$ & 2.73 & $0.16 \pm 0.02$ \\
  p-c-C$_3$H$_2$ ($2_{2,0}-1_{1,1}$) & $-13.2$ & $3.2 \pm 0.7$ & $0.025 \pm 0.004$ & $0.007 \pm 0.004$ & 2.73 & $0.25 \pm 0.04$ \\
  p-c-C$_3$H$_2$ ($2_{2,0}-1_{1,1}$) & $-10.3$ & $0.60 \pm 0.05$ & $0.032 \pm 0.002$ & $0.051 \pm 0.004$ & 2.73 & $0.32 \pm 0.02$ \\
  p-c-C$_3$H$_2$ ($2_{2,0}-1_{1,1}$) & $-8.5$ & $1.3 \pm 0.1$ & $0.042 \pm 0.003$ & $0.031 \pm 0.004$ & 2.73 & $0.42 \pm 0.03$ \\
   p-c-C$_3$H$_2$ ($2_{2,0}-1_{1,1}$) & $-4.0$ & $1.97 \pm 0.23$ & $0.032 \pm 0.003$ & $0.015 \pm 0.004$ & 2.73 & $0.32 \pm 0.03$ \\
 \hline
\end{tabular}
   \tablefoot{\tablefoottext{a}{The column densities are computed using the factors listed in Table \ref{tab:lines} when $T_{ex}=2.73$~K.       } 
           }
    \label{tab:fits}
  \end{table*}
  }

  \newcommand{\Tabcol}{
  \begin{table}
    \centering %
    \caption{References for rate coefficients.}
    \begin{tabular}{lll}
  \hline \hline
  Species & collision & Reference \\
\hline
CS & ortho H$_2$ & \citet{Denis:2018} \\
CS & para H$_2$ & \citet{Denis:2018}  \\
CS & He & \citet{Lique:2007}\\
CS & e$^-$ & \citet{Varambhia}\\
  \hline
H$_2$CO & ortho H$_2$ & \citet{Wiesenfeld} \\  
 H$_2$CO & para H$_2$ & \citet{Wiesenfeld}  \\  
 H$_2$CO & He & \citet{Green:1991}\\  
  H$_2$CO & e$^-$ & This work, see text\\    
 \hline
c-C$_3$H$_2$ & para H$_2$ & \citet{Khalifa:2019}\\   
c-C$_3$H$_2$ & He & \citet{Khalifa:2019}\\    
 \hline
  \end{tabular} 
  \tablefoot{Rate coefficients have been extracted from the BASECOL, EMAA, and LAMDA data bases.
  
  }
    \label{tab:cols}
  \end{table}
  }

    \newcommand{\Tabab}{
  \begin{table*}
    \centering %
    \caption{Column densities, abundances, and ortho-to-para ratios.}
    \begin{tabular}{lccccc}
  \hline \hline
 Velocity (\kms)                                     & $-17$                     & $-10$                  & $-13$                    & $-8$                     & $-4$  \\
\hline
N(H$_2$)$^a$ (10$^{20}$ cm$^{-2}$) & $4.18 \pm 0.1$        & $4.26 \pm 0.1$    & $5.07 \pm 0.15$  &  $3.96 \pm 0.15$  &  $3.89 \pm 0.3$  \\
N(CO)$^b$ (10$^{15}$ cm$^{-2}$)      & $6.6 \pm 0.5$           & $4.4 \pm 0.2$     & $1.0 \pm 0.07$    & $2.6 \pm 0.08$     &  $1.3 \pm 0.07$    \\
N($^{13}$CO$)^b$ (10$^{14}$ cm$^{-2}$) &  $4.3 \pm 0.5$   & $1.8 \pm 0.3$     & $< 0.35$             &  $< 0.56$               & $0.41 \pm 0.16$  \\

[CO]$^b$ (10$^{-5}$)                                  &  $ 1.56 \pm 0.15$  & $1.03 \pm 0.2 $     & $0.2 \pm 0.02$ & $0.66 \pm 0.03$	   & $0.33\pm 0.02$ \\
\hline
 [CS] ($10^{-9}$)                                    & $7.9 \pm 0.7 $           &     $15 \pm 0.7 $  &   $<0.7$        & $3.3 \pm 1.0 $        & $1.3 \pm 0.8 $  \\

 [C$^{34}$S]  ($10^{-10}$) &                  $3.1 \pm 0.7 $            &      $7.9 \pm 0.8$     & ...                   & ...                        & ...  \\
\hline
[o-H$_2$CO] ($10^{-9}$)  & $5.3 \pm 0.2$  & $8.2 \pm 0.5$ & $1.7 \pm 0.1$ &$3.5 \pm 0.3$ & $2.2 \pm 0.2$  \\

 [p-H$_2$CO]  ($10^{-9}$) & $1.5 \pm 0.2$ & $2.6 \pm 0.2$ & $0.55 \pm 0.1$ & $1.2 \pm 0.2$ & $0.72 \pm 0.2$  \\

[o-H$_2$$^{13}$CO]  ($10^{-10}$) & ... &$1.3 \pm 0.2$ & ...& ...& ...\\
\hline
o/p \HHCO & $3.5 \pm 0.5$ & $3.2 \pm 0.4$ & $3 \pm 0.6$ & $2.9 \pm 0.5$ & $3.0 \pm 0.6$ \\
\hline
[o-c-C$_3$H$_2$] ($10^{-10}$) & $8.6 \pm 2$ & $23 \pm 2 $ & $9.3 \pm 2 $ & $32 \pm 5 $ & $18 \pm 4$ \\

 [p-c-C$_3$H$_2$]  ($10^{-10}$)  & $3.8 \pm 0.5$ & $7.5 \pm 0.5$ & $4.9 \pm 0.8$ & $11 \pm 0.8 $ & $8.2 \pm 0.8$ \\
 
\hline
o/p c-C$_3$H$_2$  & $2.2 \pm 0.7$ & $3.0 \pm 0.5$ & $1.9 \pm 0.7$ & $3.0 \pm0.7$ & $2.2 \pm0.6$ \\
 \hline
 
 [CCH]$^c$ (10$^{-9}$)                    &     28.1	& 53.2  & 29.2  &	60.2      & 46.1 \\
 
 [HCN]$^d$ (10$^{-9}$)                  & 7.03  &  8.50 &  1.78        & 4.32 & 	3.18 \\
 \hline
  \end{tabular} 
  \tablefoot{
  \tablefoottext{a}{From HCO$^+$ and assuming an abundance of $3\times 10^{-9}$}
  \tablefoottext{b}{From \citet{ll98}}
   \tablefoottext{c}{From \citet{ll00}}
    \tablefoottext{d}{From \citet{ll01}}
  }
    \label{tab:abund}
  \end{table*}
  }

      \newcommand{\Tabratio}{
  \begin{table}
  \caption{Ratios of integrated optical depth to the integrated optical depth of the -10 \kms \, velocity component.}
  \begin{tabular}{lccccc}
  \hline
  \hline
Velocity (\kms) & $-17$ & $-14$ & $-10$ &  $-8$ & $-4$ \\ 
  \hline
  o-H$_2$CO (150 GHz) & 1/1.56&	1/3.8 &	1/1& 	1/2.2 & 	1/3.8 \\
  p-H$_2$CO (145 GHz) &1/1.72&	1/3.8 &	1/1&	1/2.1	& 1/3.8 \\
  o-H$_2$CO (4.8 GHz) & 1/1.95 &	1/7.8	 & 1/1	& 1/3.0	& 1/6.4 \\
  \hline
  \end{tabular}
    \tablefoot{The intrinsic N(H$_2$CO) ratios should follow the mm-lines, and if the ratio $\tau(v=x)/\tau(v=-10)$ is smaller in the 4.8 GHz line, that should indicate higher electron
    fraction in the $v=x$ component than in the $-10$~\kms \, component.
}
 \label{tab:ratio}
 \end{table}
  }
  
      \newcommand{\Tabref}{
  \begin{table}
 \caption{\rev{References for the ancillary data.}}
  \begin{tabular}{lcl}
  \hline
  \hline
Species  & Line & Reference \\ 
  \hline
  o-H$_2$CO  & $1_{1,0}-1_{1,1} $ & \citet{ll95} \\ 
  CS                                &   $2-1$                   &\citet{ll02} \\ 
  o-c-C$_3$H$_2$             & $2_{1,2}-1_{0,1} $ & \citet{ll00} \\ 
   o-c-C$_3$H$_2$             & $1_{1,0}-1_{0,1} $ & \citet{liszt:2012} \\ 
  HCO$^+$ &                   $1-0$                   & \citet{ll96-1} \\
  CCH                       &     $1_{3/2,2}-0_{1/2,1}$            & \citet{ll00}\\
  \hline
  \end{tabular}
 \label{tab:ref}
 \end{table}
  }

\newcommand{\Figlines}{%
  \begin{figure*}
    \centering %
    \includegraphics[width=0.85\linewidth]{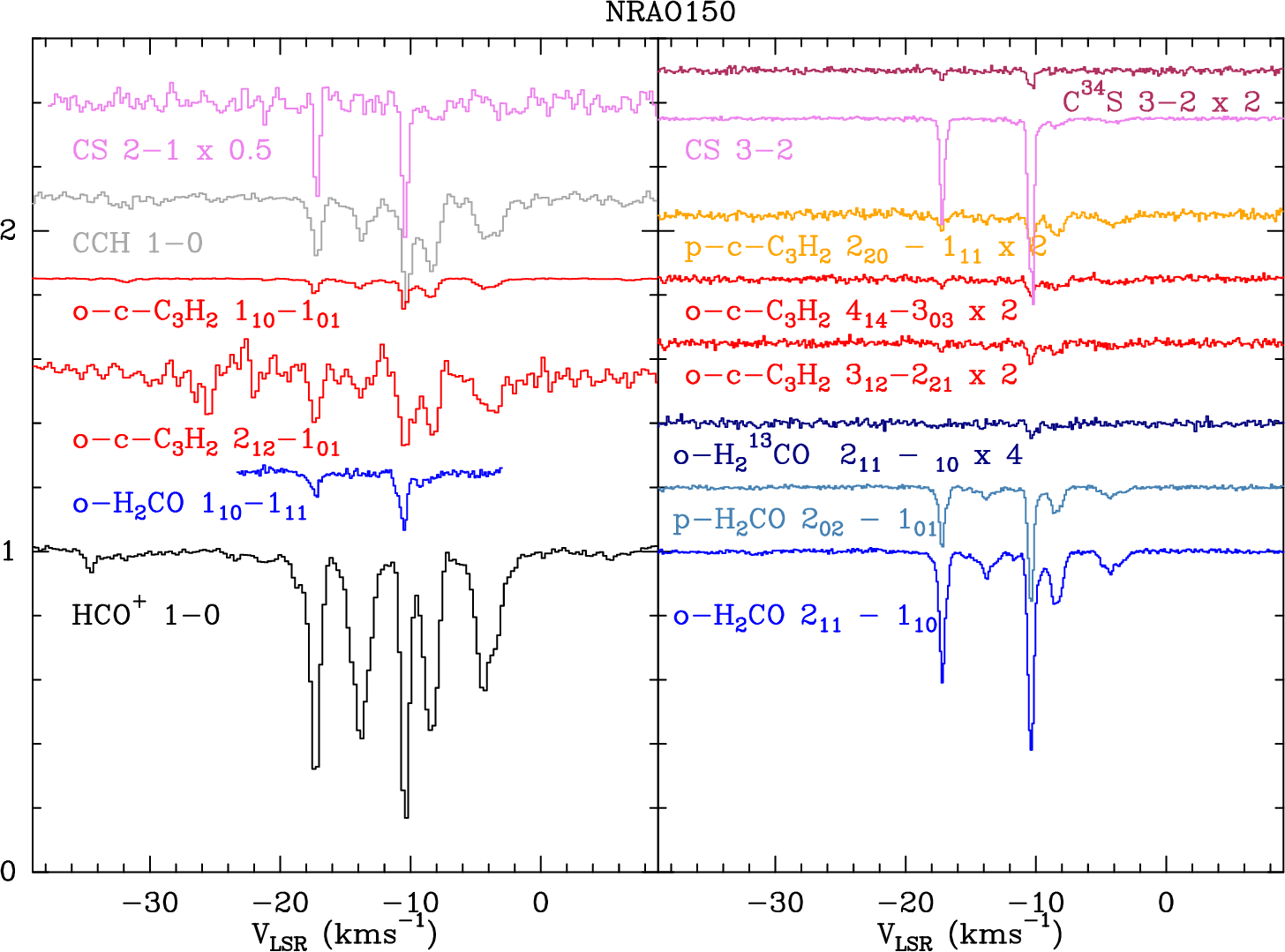}
    \caption{Absorption lines detected toward NRAO 150. The left panel shows previous detections including the o-H$_2$CO ($1_{1,0}-1_{1,1}$)  line at 4.8~GHz and
    the  o-c-C$_3$H$_2$ ($1_{1,0}-1_{0,1}$) line at 18~GHz observed at the VLA \citep{ll95,liszt:2012} , and the  CS (2-1) (98~GHz), o-c-C$_3$H$_2$ ($2_{1,2}-1_{0,1}$)(85.3~GHz), CCH($1,3/2,2 - 0,1/2,1$) (87.3~GHz), and HCO$^+$($1-0$) (89.2~GHz) lines \citep{ll94,ll95,ll96-1,ll00,ll02} observed  with the PdBI. The right panel presents the new observations including the
  o-H$_2$CO($2_{1,1}-1_{0,2}$) , o-H$_2$$^{13}$CO($2_{1,1}-1_{0,2}$), and p-H$_2$CO ($2_{0,2}-1_{0,1}$) lines, the   CS(3-2), C$^{34}$S(3-2) lines, and the
    c-C$_3$H$_2$ transitions listed in Table \ref{tab:lines}.  All displayed spectra have been  normalized by the continuum strength. Vertical offsets have been used for clarity. }
    \label{fig:lines}
  \end{figure*}}
  
 \newcommand{\FigCS}{%
  \begin{figure}
    \includegraphics[width=0.9\linewidth]{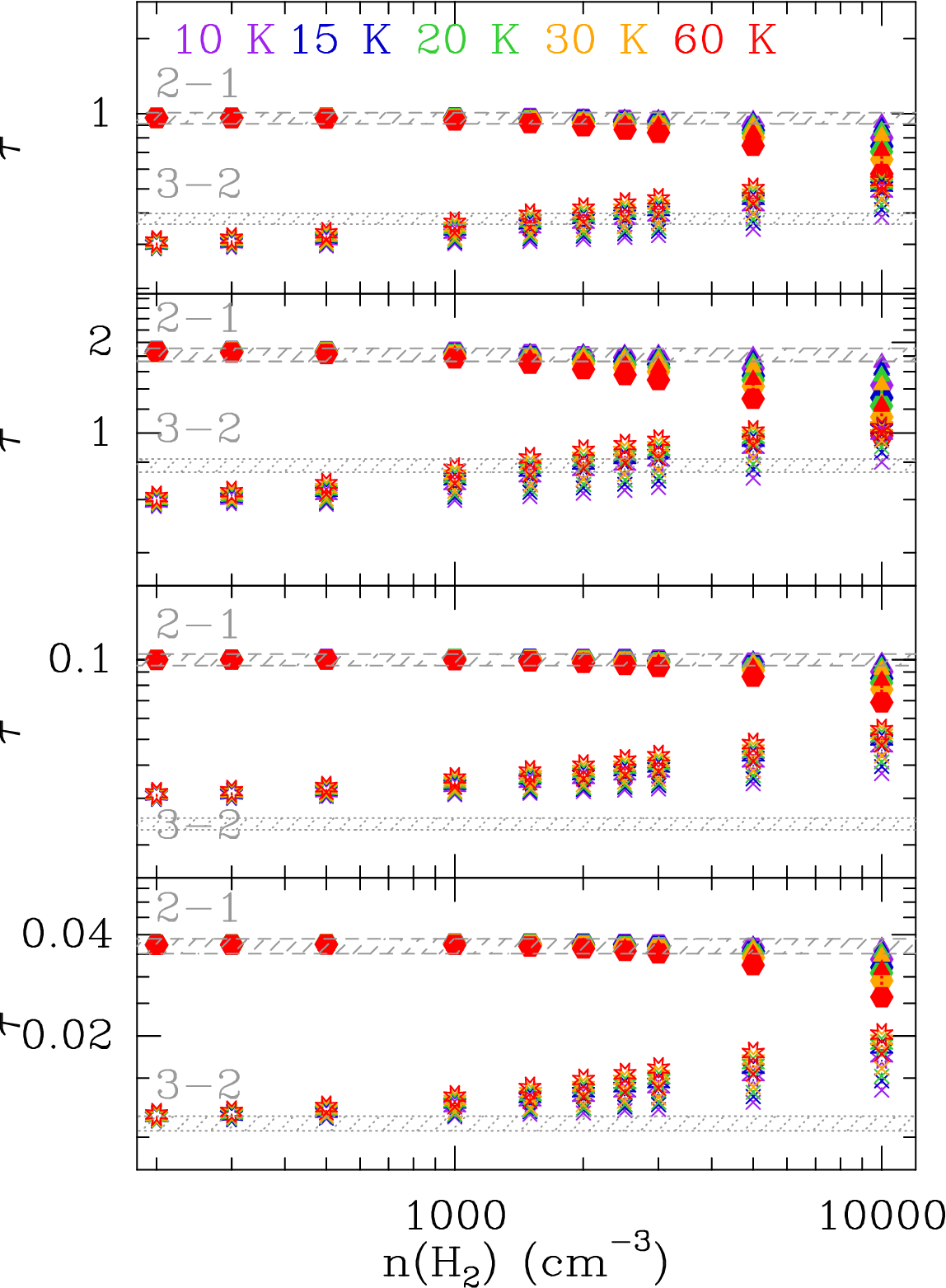}
    \caption{Variation of the predicted opacities of the CS($2-1$)  (filled symbols) and CS$(3-2)$  (starred symbols)  transitions as a function of the molecular hydrogen density in the diffuse and translucent cloud conditions computed with RADEX
 using the column densities and FWHM  listed in Table \ref{tab:fits}.
   The colors show the kinetic temperature from 10 to 60~K, and the H$_2$  ortho-to-para ratio scales with the kinetic temperature.  The symbol shapes indicate the electron fraction as follows for the CS($2-1$) line: lower than  $2\times 10^{-5}$ filled triangles, between $2\times 10^{-5}$ and 10$^{-4}$, small filled squares, and 
  filled circles for $2\times 10^{-4}$. A similar progression is used for the CS($3-2$) line, lower than  $2\times 10^{-5}$ crosses, between $2\times 10^{-5}$ and 10$^{-4}$, small stars, and  8-branch stars for $2\times 10^{-4}$. 
The measured opacities for the four detected velocity components are shown with the dashed and dotted gray areas, from top to bottom: $-17.2$~\kms , $-10.3$~\kms ,  $-8.5$~\kms \, and $-4$~\kms .    }
    \label{fig:cs}
  \end{figure}}
  
 \newcommand{\FigH}{%
  \begin{figure*}
   \centering %
    \includegraphics[width=0.4\linewidth]{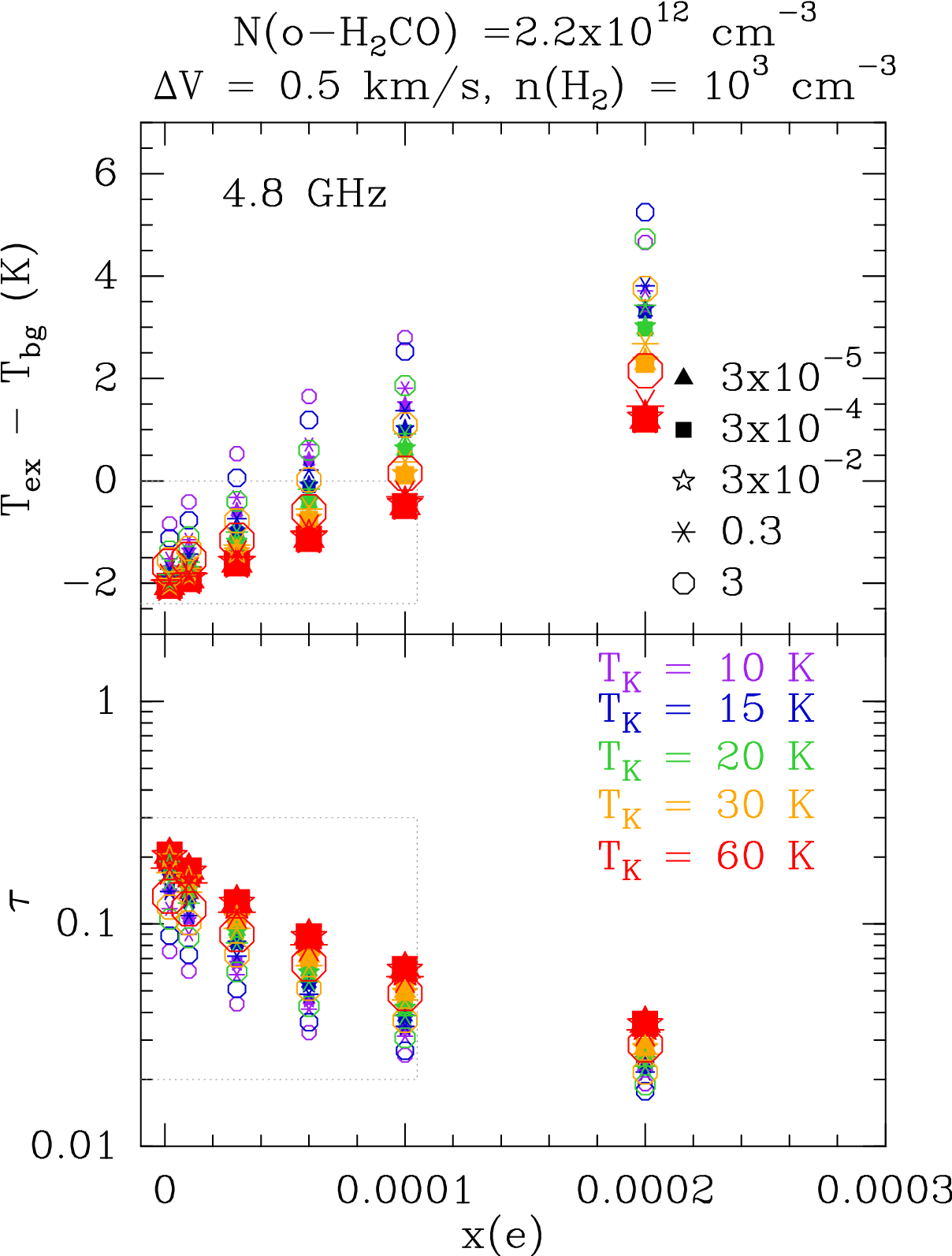}
     \includegraphics[width=0.4\linewidth]{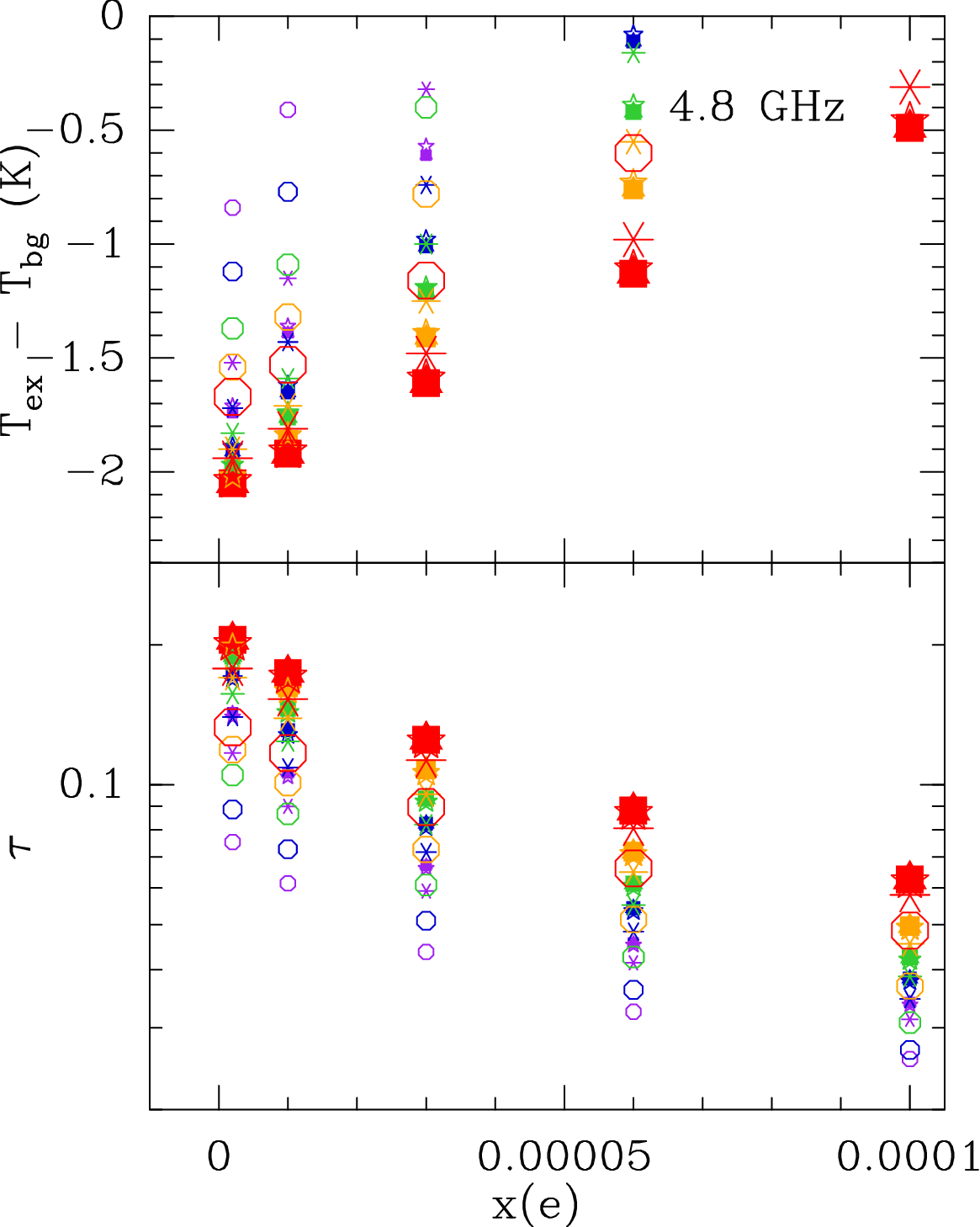}
      \includegraphics[width=0.4\linewidth]{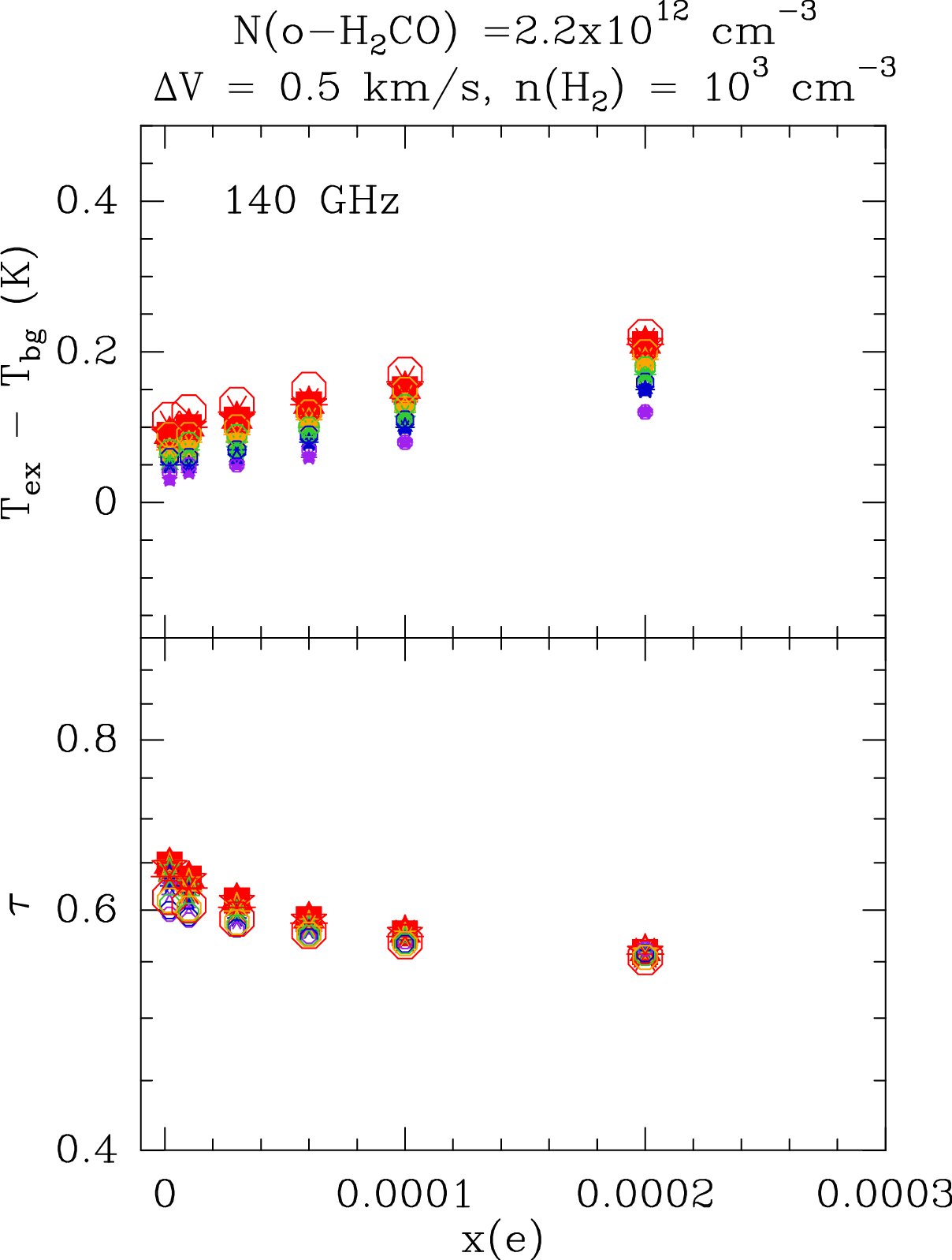}
           \includegraphics[width=0.4\linewidth]{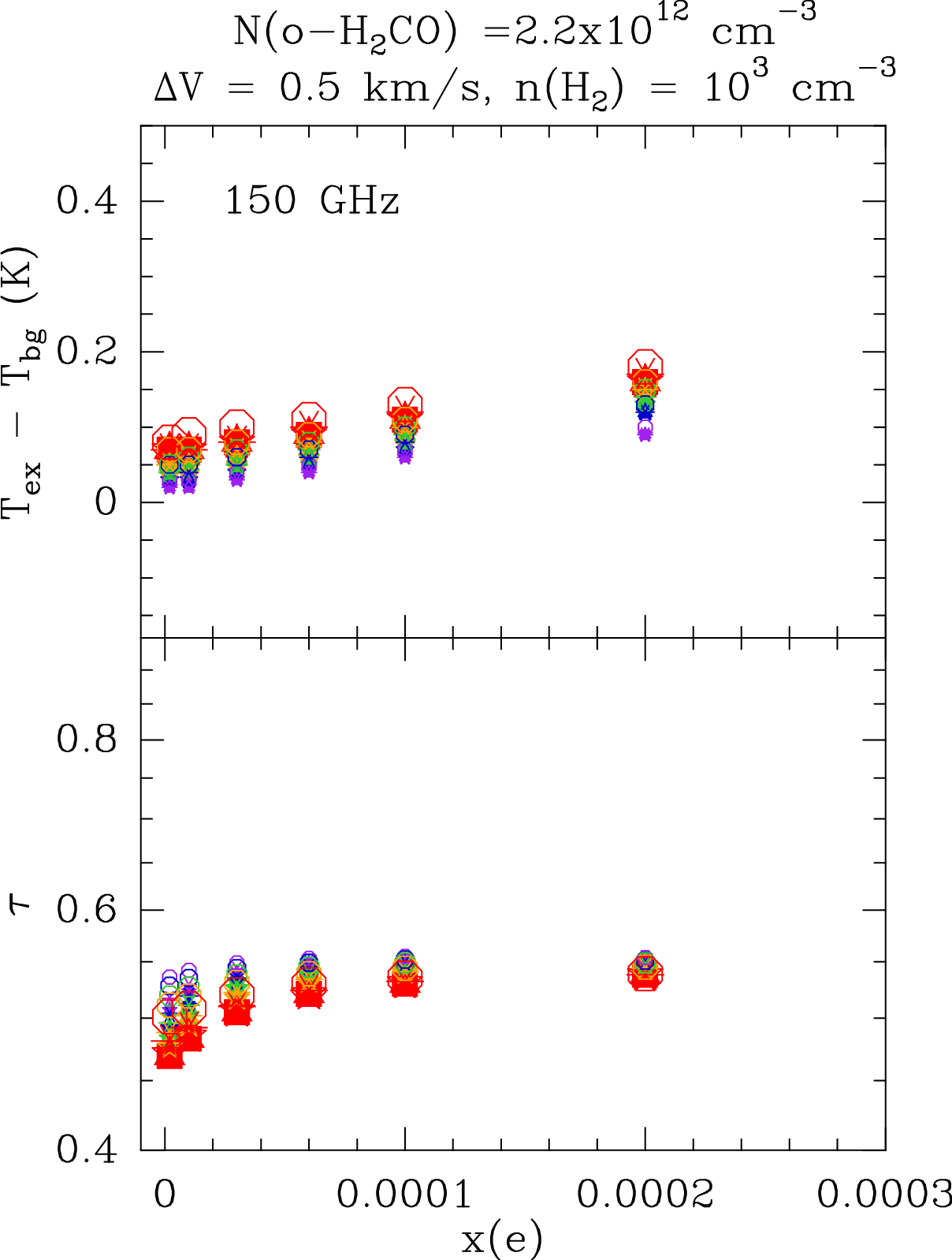}
    \caption{Variation of the excitation temperature and opacity with the electron  fraction, kinetic temperature, and H$_2$ ortho-to-para ratio
    for the $1_{1,0}-1_{1,1}$ line at 4.8~GHz  (top left), and zoom on the region with excitation temperature lower than the CMB indicated by a dotted box (top right). 
    The $2_{1,2}-1_{1,1}$ line at 140~GHz is shown in the bottom left panel and the $2_{1,1}-1_{1,0}$ line at 150~GHz in the bottom right panel.  The difference between the excitation temperature and the CMB is plotted to illustrate the change of regime of the $1_{1,0}-1_{1,1}$ transition at 4.8~GHz, and a different scale is used for the 140~GHz and 150~GHz lines.  Different symbols show the values of the H$_2$ ortho-to-para ratio between $3\times 10^{-5}$ and 3, their color and size indicate the  kinetic temperature between 10\,K and 60\,K. The column density of o-H$_2$CO is $2.2\times 10^{12}$\, cm$^{-2}$, 
the H$_2$ density is set to 10$^3$ cm$^{-3}$ and the FWHM is 0.5 \kms. The $2_{1,2}-1_{1,1}$ line at 140~GHz has not been observed in this work. }
    \label{fig:h2co}
  \end{figure*}}

 \newcommand{\FigPH}{%
  \begin{figure*}
   \centering %
    \includegraphics[width=0.3\linewidth]{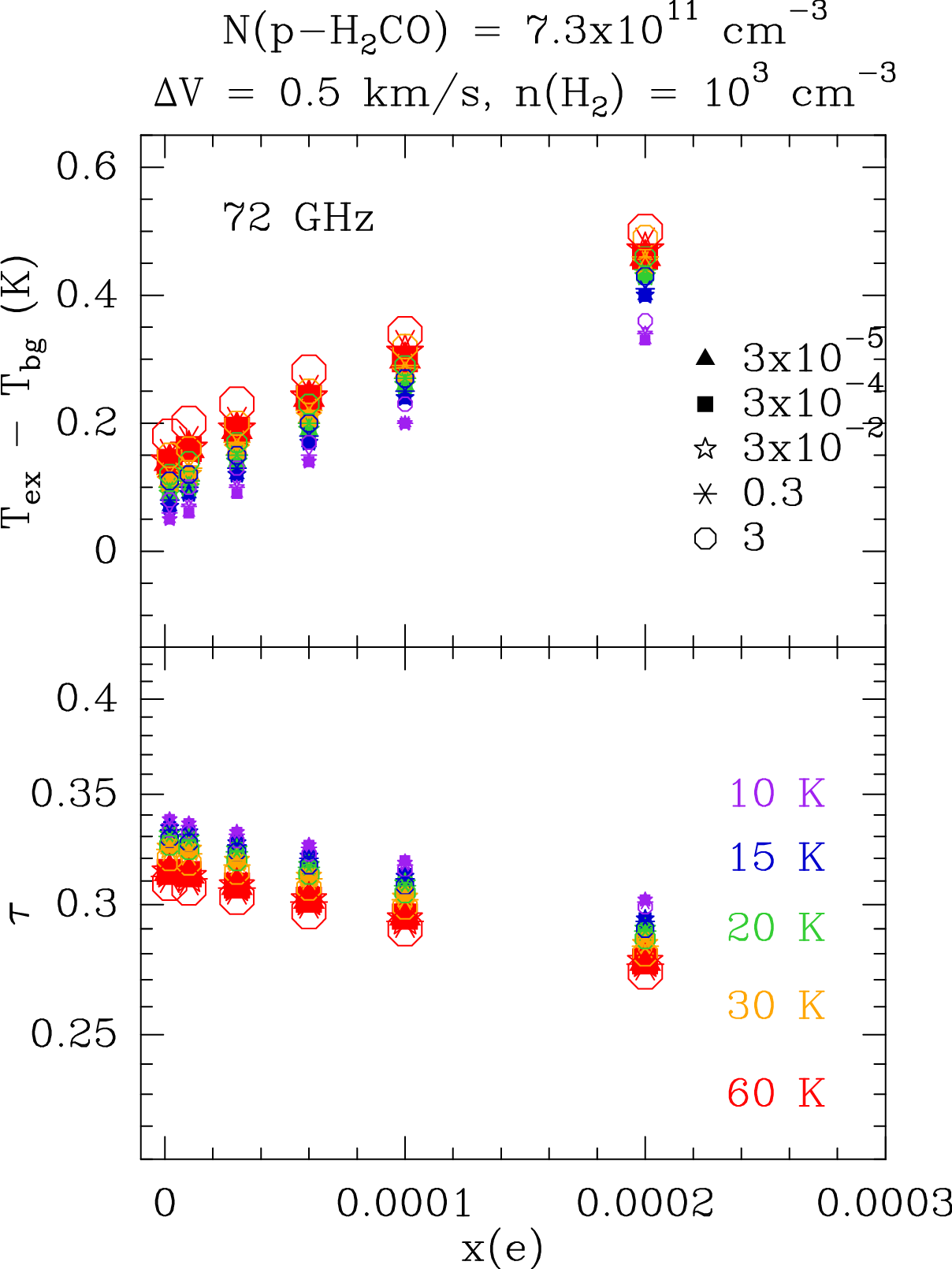}
      \includegraphics[width=0.3\linewidth]{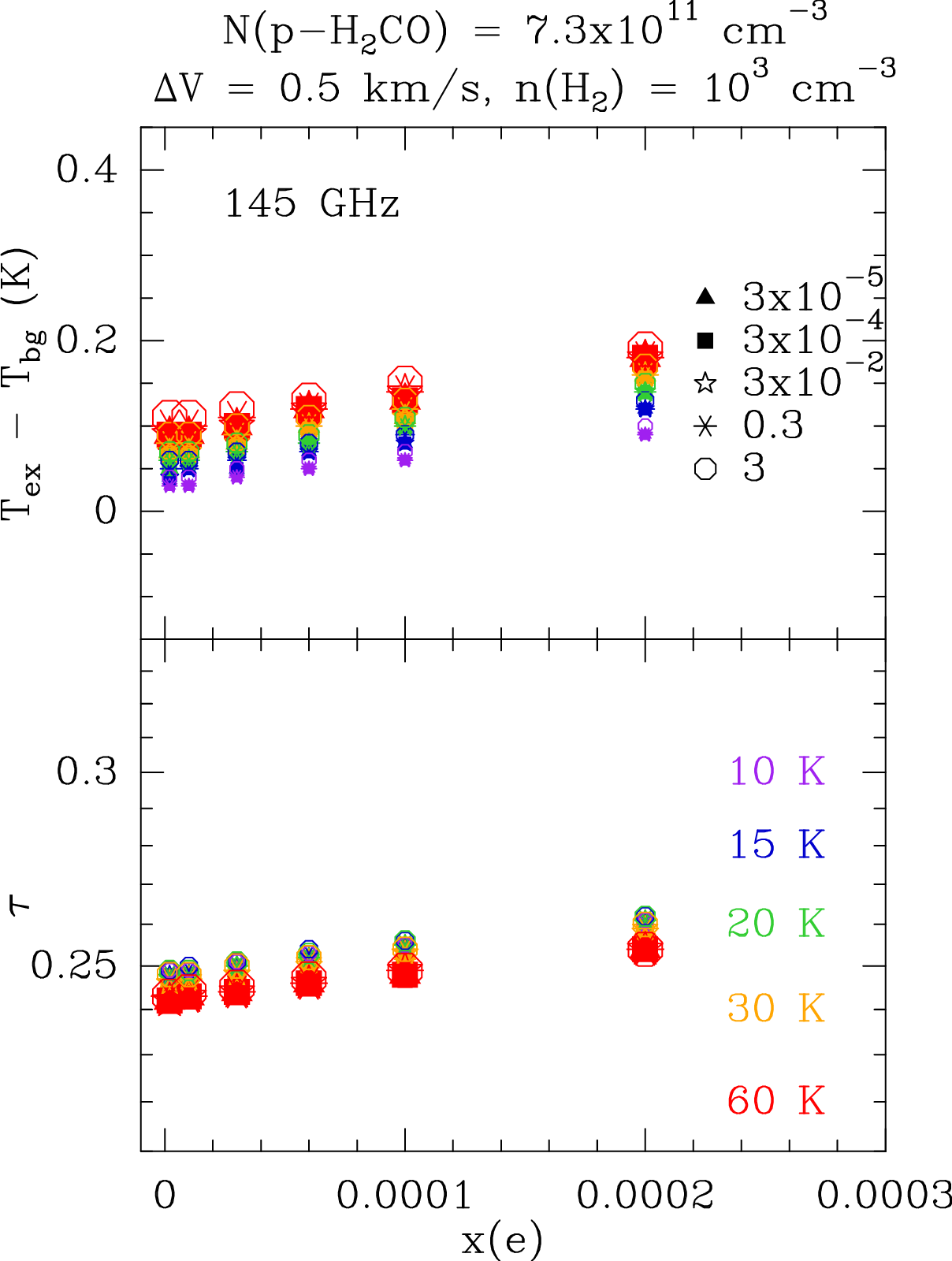}
    \caption{Variation of the excitation temperature and opacity for the $1_{0,1}-0_{0,0}$ line at 72~GHz  (left), and the $2_{0,2}-1_{0,1}$ line at 145~GHz (right) with the electron  fraction, kinetic temperature and H$_2$ ortho-to-para ratio. The difference between the excitation temperature and the CMB is plotted. Different symbols show the values of the H$_2$ ortho-to-para ratio between $3\times 10^{-5}$ and 3, their color and size indicate the  kinetic temperature between 10\,K and 60\,K. The column density of p-H$_2$CO is $7.3\times 10^{11}$\, cm$^{-2}$,  the H$_2$ density is set to 10$^3$ cm$^{-3}$, and the line FWHM is 0.5 \kms. The $1_{0,1}-0_{0,0}$ line at 72~GHz has
    not been observed in this work.  }
    \label{fig:ph2co}
  \end{figure*}}
  
\newcommand{\FigGRO}{%
  \begin{figure*}
   \centering %
    \includegraphics[width=0.3\linewidth]{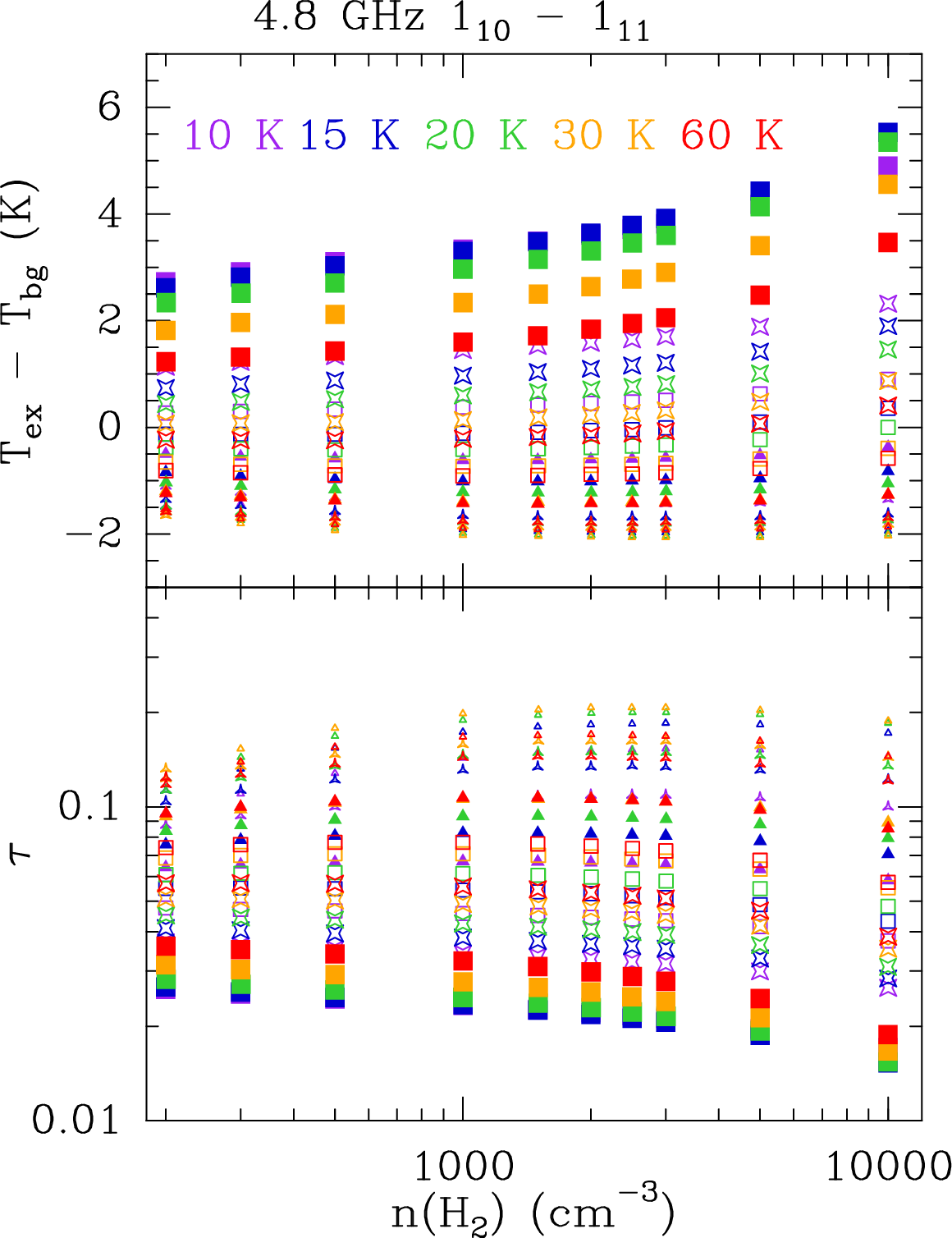}
      \includegraphics[width=0.3\linewidth]{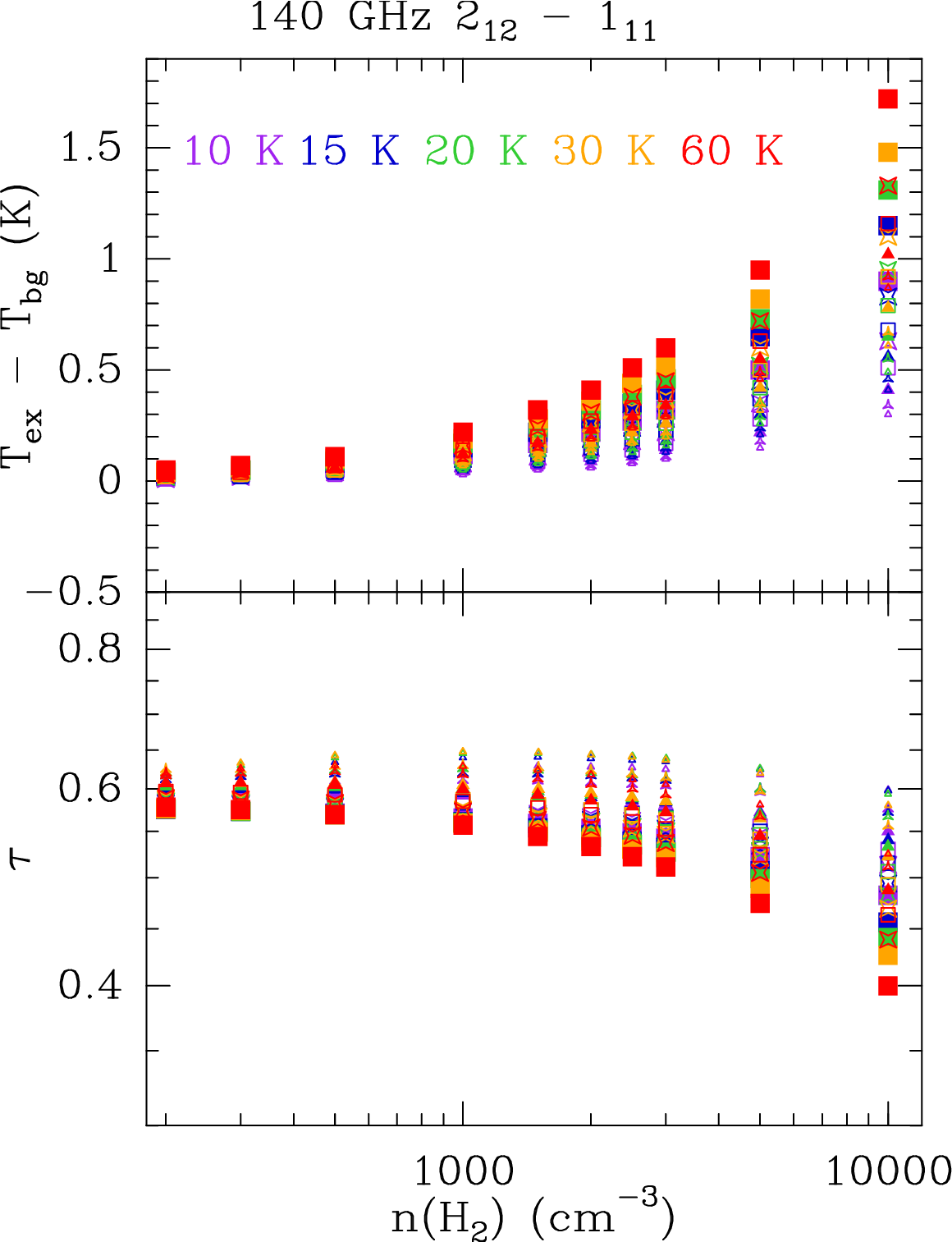}
        \includegraphics[width=0.3\linewidth]{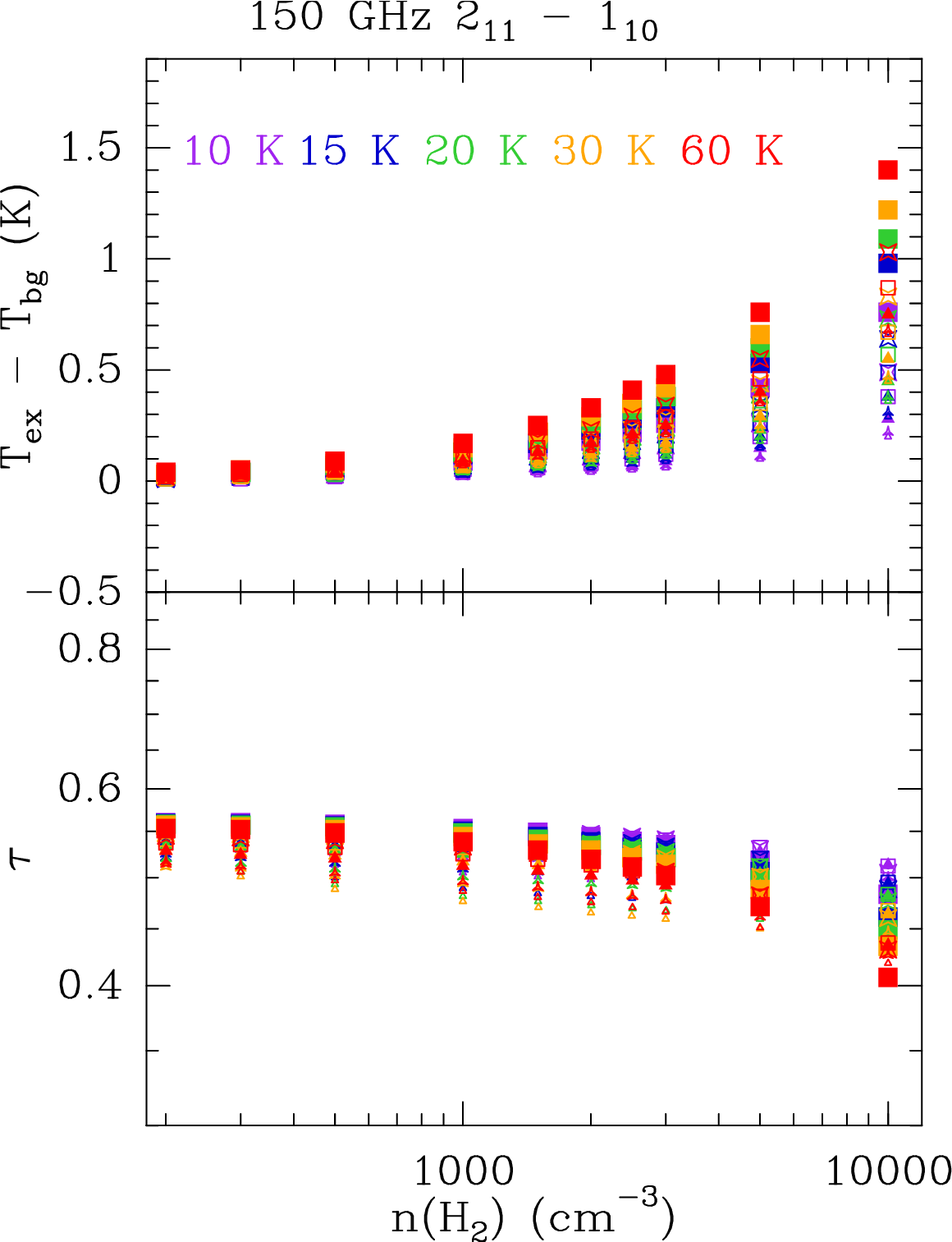}
    \caption{Variation of the predicted excitation temperature (top) and opacity (bottom) for the three lowest lines of o-H$_2$CO. The left panel 
    shows the $1_{1,0}-1_{1,1}$ line at
     4.8~GHz, the middle panel shows the $2_{1,2}-1_{1,1}$ line at 140~GHz, and the right panel the  $2_{1,1}-1_{1,1}$ line at 150~GHz. The o-H$_2$CO column density is set to $2.2\times 10^{12}$~cm$^{-2}$, the line FWHM is 0.5\kms. The kinetic temperature varies from 10\,K to 60\,K. The H$_2$ ortho-to-para ratio increases with the
     kinetic temperature. The symbol sizes and shapes indicate the electron fraction : empty triangles for $2\times10^{-6}$, starred triangles for $10^{-5}$, filled triangles for $3\times10^{-5}$, empty squares for $6\times10^{-5}$, starred squares for $10^{-4}$, and filled squares for  $2\times10^{-4}$. Larger symbols are used for larger values of the electron fraction. The $2_{1,2}-1_{1,1}$ line at 140~GHz has not been observed in this work.
     }
    \label{fig:goh2co}
  \end{figure*}}
  
  \newcommand{\FigGRP}{%
  \begin{figure}[h!]
   \centering %
      \includegraphics[width=0.8\linewidth]{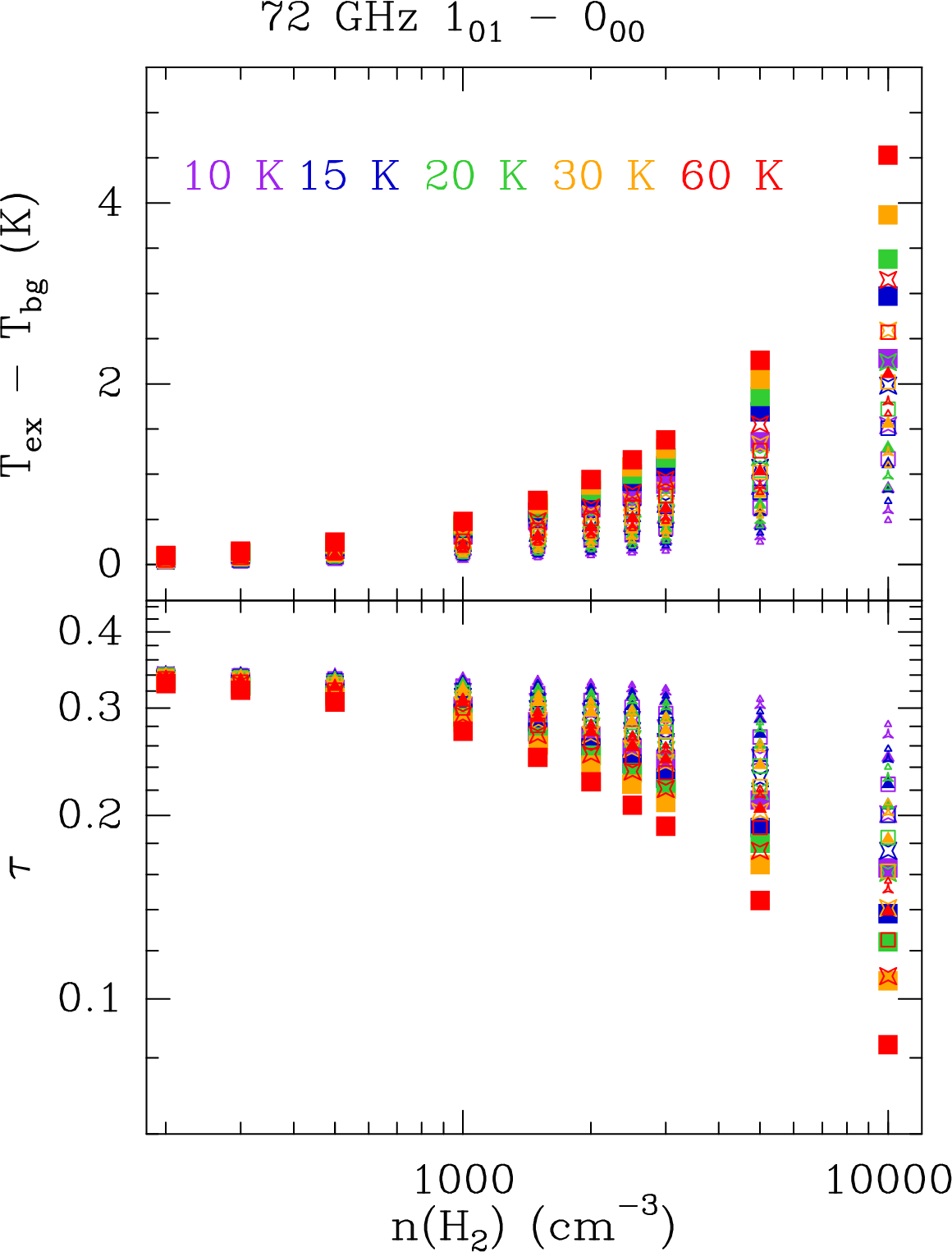}
      \includegraphics[width=0.8\linewidth]{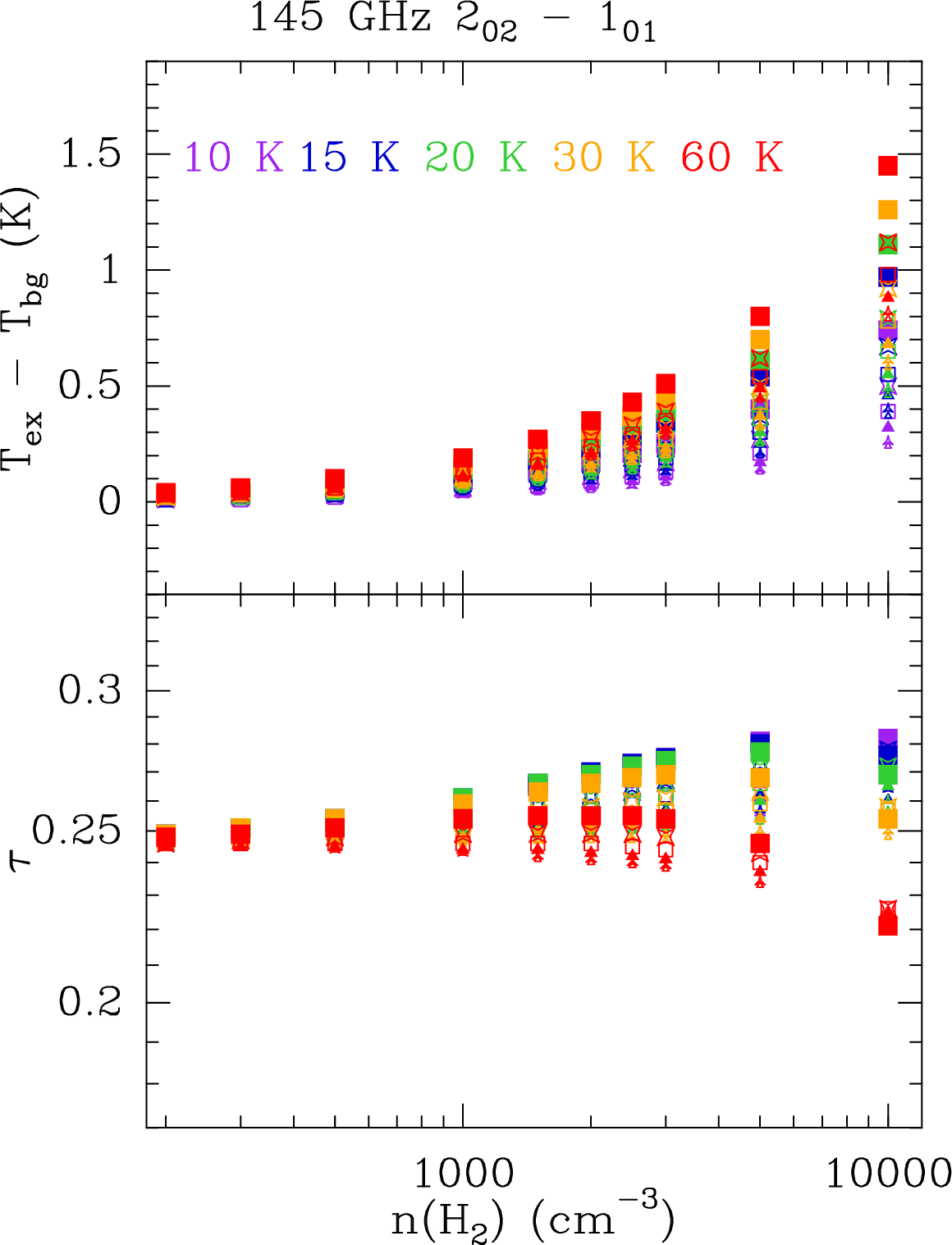}
    \caption{
    Variation of the predicted excitation temperature (top) and opacity (bottom) for the two lowest lines of p-H$_2$CO. The left panel shows the $1_{0,1}-0_{0,0}$ line at
     72~GHz, and the right panel shows the $2_{0,2}-1_{0,1}$ line at 145~GHz. The p-H$_2$CO column density is set to $7.3\times 10^{11}$~cm$^{-2}$, the line FWHM is 0.5\kms. The kinetic temperature varies from 10\,K to 60\,K. The H$_2$ ortho-to-para ratio varies with the
     kinetic temperature. The symbol sizes and shapes indicate the electron fraction : empty triangles for $2\times10^{-6}$, starred triangles for $10^{-5}$, filled triangles for $3\times10^{-5}$, empty squares for $6\times10^{-5}$, starred squares for $10^{-4}$, and filled squares for  $2\times10^{-4}$. Larger symbols are used for larger values of the electron fraction. The $1_{0,1}-0_{0,0}$ line at 72~GHz has
    not been observed in this work.  
       }
    \label{fig:gph2co}
  \end{figure}}
  
  \newcommand{\FigC}{%
  \begin{figure*}
   \centering %
      \includegraphics[width=0.8\linewidth]{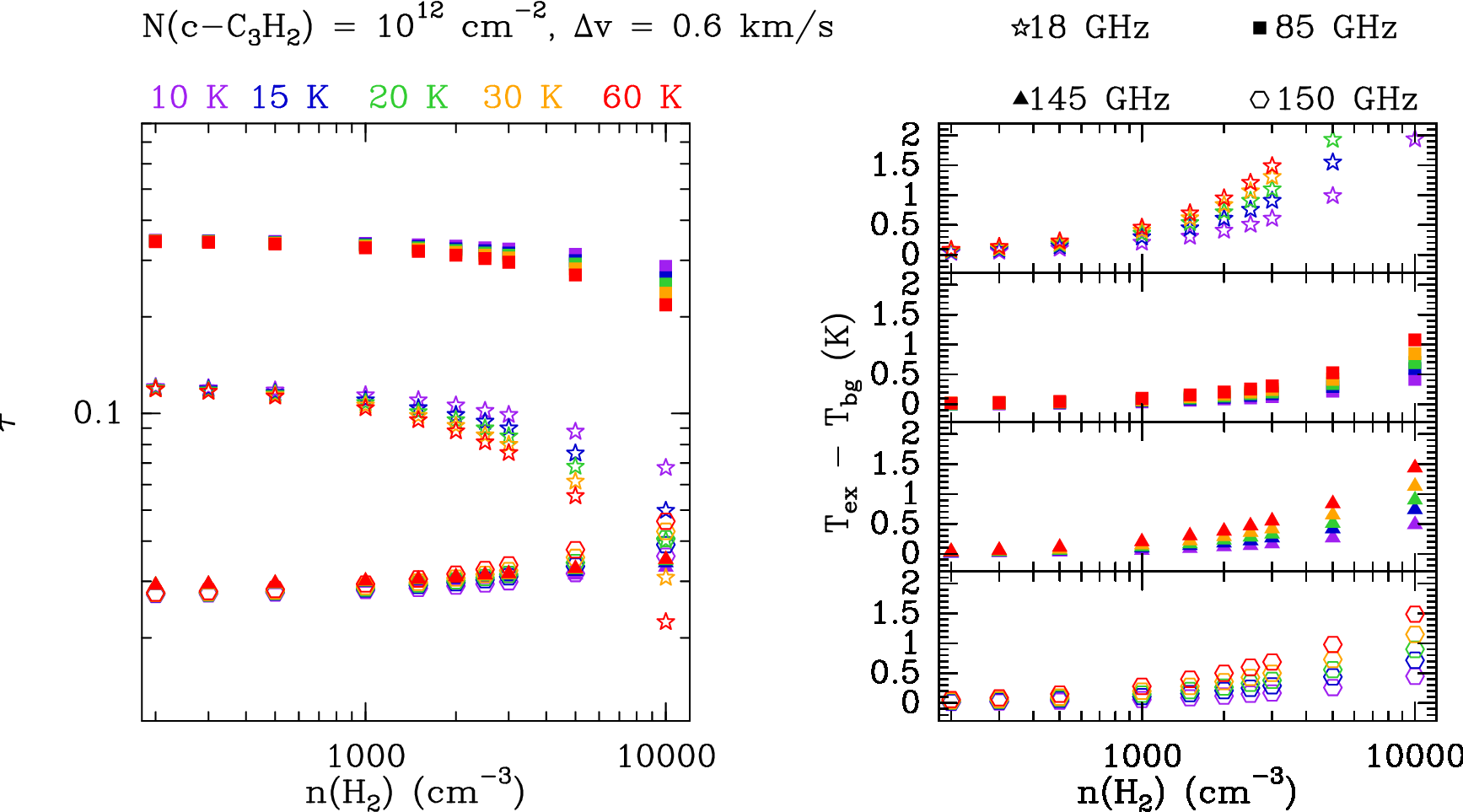}
   
    \caption{Variation of the opacity and excitation temperature of the four detected lines of ortho c-C$_3$H$_3$, as a function of the H$_2$ density and kinetic temperature.
    The calculations used  collision cross sections with He from \citet{Khalifa:2019}, and their scaled version for para H$_2$. Stars show the
    values for the $1_{1,0}-1_{0,1}$ line at 18~GHz, filled squares for the $2_{1,2}-1_{0,1}$ line at 85~GHz, filled triangles for the $3_{1,2}-2_{2,1}$ line at 145~GHz, and
    empty circles for the $4_{1,4}-3_{0,3}$ line at 150~GHz. The left panel presents the line opacities and the right panel the difference between the  excitation temperature
    and the CMB. The adopted column density is 10$^{12}$ cm$^{-2}$ and the line FWHM is 0.6 kms$^{-1}$. 
       }
    \label{fig:c3h2}
  \end{figure*}}

\begin{document} 

\title{H$_2$CO and CS in diffuse clouds: Excitation and abundance}

\author{
  Maryvonne Gerin\inst{1} %
    \and Harvey Liszt\inst{2} %
  \and Jérôme Pety\inst{3,1} %
  \and Alexandre Faure\inst{4}%
  }

\institute{LERMA, Observatoire de Paris, PSL Research University, CNRS,
  Sorbonne Université, 75014 Paris, France 
   \and National Radio Astronomy Observatory, 520 Edgemont Road,
  Charlottesville, VA, 22903, USA 
   \and IRAM, 300 rue de la Piscine, 38406 Saint Martin d’Hères, France 
   \and   Univ. Grenoble Alpes, CNRS, IPAG, 38000 Grenoble, France.
}

\date{Received January 2024; accepted yy   2024}

 
\abstract
{Diffuse interstellar clouds present an active chemistry despite their relatively low density and the ubiquitous presence of far-UV radiation. 
}
{To provide constraints on the chemical processes responsible for the observed columns of  organic species, we used the NOEMA interferometer to observe the sight line toward NRAO150 (B0355+508)  in the 2mm spectral window.}
{We targeted the low excitation lines of ortho H$_2$CO ($2_{1,1}-1_{1,0}$) and para H$_2$CO ($2_{0,2}-1_{0,1}$)  as well as the nearby transitions of CS($3-2$) and c-C$_3$H$_2$ ($3_{1,2}-2_{2,1}$), ($4_{1,4}-3_{0,3}$), and ($2_{2,0}-1_{1,1}$). We combined these data with previous observations of the same sight line to determine the excitation conditions, column densities, and abundances relative to H$_2$ in the different velocity components. 
We performed non-LTE radiative transfer calculations including collision cross sections with ortho and para H$_2$ and with electrons. New collision cross sections with electrons were computed for ortho and para formaldehyde. }
{All targeted lines were detected. The c-C$_3$H$_2$ line profiles are very similar to those of HCO$^+$ and CCH, while the CS absorption features are narrower and mostly concentrated in  two main velocity components at V$_{LSR}$ = -17.2 and -10.4 \kms. H$_2$CO absorption lines present an intermediate pattern with absorption in all velocity components but  larger opacities in the two main velocity components. The ortho-to-para ratios of H$_2$CO and c-C$_3$H$_2$ are consistent with the statistical value of three. While the excitation temperature of all c-C$_3$H$_2$ velocity components is consistent with the Cosmic Microwave Background (CMB), the two strong components detected in CS show a clear excess over the CMB indicating that CS resides at higher densities than other species along this particular sightline, n(H$_2$) $\sim 2500$~ cm$^{-3}$ while n(H$_2$) $< 500$~ cm$^{-3}$ for the other velocity components. We detected faint absorption from o-H$_2$$^{13}$CO and C$^{34}$S allowing us to derive isotopic ratios: o-H$_2$CO/o-H$_2$$^{13}$CO = $61\pm 12$ and  C$^{32}$S/C$^{34}$S = $24 \pm6$. 
 The excitation of the $1_{1,0}-1_{1,1}$ line of formaldehyde at 4.8~GHz is sensitive to the electron fraction and  its excitation temperature is predicted to be lower than the CMB at low and moderate electron fractions ($x(e) < 6\times 10^{-5}$), and to rise above the CMB at high electron fractions ($x(e) >  10^{-4}$).  }
{}


\maketitle{} %

\section{Introduction}

The diffuse interstellar medium is known to host a rich polyatomic
chemistry despite the low densities, of a few tens to a few hundreds of molecules per cubic centimeter \citep{snow:06}, 
and the ubiquitous illumination by far-UV radiation.The moderate reddening
E(B-V)$\sim 0.1\,$mag at which milllimeter-wave absorption is indeed observed from \HCOp{}
and other species, just above the threshold for \Ht{} formation, \citep{Liszt:2018a,Liszt:2018b}
ensures that UV radiation permeates the material, making photodissociation the
main destruction routes for neutral molecules.  In those conditions, carbon is largely ionized,
consistent with the small observed CO and C\textsc{I} column densities \citep{Liszt2023}. Therefore, the
electron fraction is high and the gas is warm, with mean kinetic
temperatures $\sim 75\,$K determined in \Ht{} at UV wavelengths \citep{Shull:21}, and decreasing to $\sim 35 - 50$K in the slightly denser 
material absorbing in C$_2$ \citep{sonnentrucker:07}. Despite
this, the relative abundances of many molecules observed in both diffuse
and cold dark molecular clouds are strikingly similar \citep{ll95}.

Given the low densities, collisional excitation is weak and the internal
energy level population is mainly controlled by the cosmic background
radiation \citep[e.g.,][]{Liszt:2016}. Absorption lines from low-lying levels are easily detected with a
high signal-to-noise against background continuum sources, whether
stars in the UV-optical domain or quasars at radio frequencies, and the
derivation of molecular column densities is straightforward. With the
  exception of CO, the weak excitation is confirmed by observing molecular
emission around even the most strongly absorbed background sources \citep{lucas:96}.

Despite having demonstrated the richness of the chemistry in diffuse
molecular gas, our understanding of it is largely conceptual, along the
lines that turbulence in the interstellar medium must somehow speed up
the formation of \HCOp{} that subsequently recombines at thermal rates to
form CO, or \HCSp{} to form CS, and presumably \chem{HCNH^+} to form CN,
HNC, and HCN \citep{godard:14,lesaffre:20,gong:2023b}. No model exists to explain, in detail, the series of polyatomic
hydrocarbons \CCH{}, \lCCCH{}, \cCCCH{}, \CCCHp{}, \lCCCHH{}, and \cCCCHH{},
or the widespread \metcy{} \citep{Liszt:2018a}

One of the most perplexing molecules in diffuse gas is indeed formaldehyde,
one of the first interstellar molecules detected at radio wavelengths.  The early \HHCO{} detection 
 along sight lines with $A_\emr{V} = 1\,$mag \citep{nash} was not predicted by gas phase models. The presence of outer envelopes of nearby dark clouds along the sight lines
was advocated by \citet{federman:91} because  active grain surface chemistry could be forming enough \HHCO. But the accumulated evidence from observations of \HHCO{} 
and other species convincingly showed that the detected absorption toward quasars is produced in the diffuse molecular gas \citep{ll95}.

\HHCO{} has two spin symmetry states,
the para states with even $K_a$ values and the  ortho states
with odd $K_a$ values and three times higher spin degeneracy. Since only reactive collisions
destroying formaldehyde can change the nuclear-spin symmetry in the gas phase,
the ortho-to-para ratio should be determined at formation and bear the mark of the responsible
chemical (gas- or solid-phase) process. Laboratory experiments have shown that the ortho/para ratio of
H$_2$O desorbed from interstellar ices is the statistical value and high-temperature local thermodynamic Equilibrium (LTE) limit
of three \citep{Hama:2018,Putaud:2019,Dupuy:2021}, which should also apply to the ortho-to-para ratio of H$_2$CO and c-C$_3$H$_2$,
 which have the same type of ortho and para symmetry states as H$_2$O.
 \citet{Yocum:2023} have indeed shown that the ortho-to-para ratio of formaldehyde that desorbed from UV-irradiated cold 
CH$_3$OH ices is consistent with the statistical value of three, and independent of the ice temperature.
Alternatively,  the gas-phase formation chemistry accounting for the nuclear-spin statistics
of H$_2$ predicts an ortho/para ratio between two and three for \HHCO (Faure et al., in preparation).
Deviations from  an ortho-to-para ratio of  three have been observed
in the interstellar medium for H$_2$CO \citep{guzman:2011} and other species
such as NH$_2$, NH$_3$, and H$_2$O \citep{Persson:2016, Faure:2013, Faure:2019}
and were attributed to nuclear-spin gas-phase formation chemistry. 


In this paper we present new $\lambda$2mm observations of ortho and para formaldehyde absorption  along the sight line toward
NRAO150, which we interpret in conjunction with an  analysis of the excitation of carbon monosulfide (CS) based on new observations of the J=3-2 line.
We also report observations of cyclopropenilydene (c-C$_3$H$_2$) lines in the 2mm spectral window.
The observations and detected species are  described in Sec. \ref{sec:obs}. Section \ref{sec:dens} presents the implication for the density structure along the sightline to NRAO150. In Section \ref{sec:col} we discuss  the derived column densities and abundances, and in Sections \ref{sec:comp} and \ref{sec:chem} the implication for the diffuse interstellar chemistry. The conclusions are  laid out in section \ref{sec:conc}.

\Tabref

\section{Observations}
\label{sec:obs}
Observations have been performed with the NOEMA interferometer in March 2022 toward NRAO150 (B0355+508). 
The channel spacing of  62.5 kHz was used, corresponding to a velocity resolution of 0.13 \kms \, at 145 GHz. 
Side band separating receivers were used allowing simultaneous observations of the lower and upper side bands. To improve the sensitivity, the same frequency was observed in both the horizontal and vertical polarization. Two frequency tunings have been observed to obtain a continuous frequency coverage between 128.4 GHz and 136.1 GHz in the lower side band and between 143.9 GHz and 151.5 GHz in the upper side band. 
Observations were performed in excellent weather conditions. The data have been processed with the GILDAS software.
The spectra were obtained at the peak of the continuum emission and normalized by the value of the continuum. The noise levels
were determined by subtracting a base line in signal-free spectral regions
and reach about 0.3\% for all spectral bands. The detected absorption features are shown in Fig. \ref{fig:lines} together with previously observed lines toward
the same source as listed in Table \ref{tab:ref}. The parameters of the newly detected lines are displayed in Table \ref{tab:lines}. 
The spectra show the previously detected velocity components. Each absorption spectrum was fitted with five Gaussian velocity components, except for CS 
where a four Gaussian component fit is enough. Table \ref{tab:fits} lists the resulting centroid velocities, Full Widths at Half Maximum (FWHM) and  opacities. 


\TabLine
\Figlines

\Tabfit

\label{sec:dens}
\FigCS

\FigH

\rev{\section{The sightline toward NRAO150}}
The sightline toward NRAO150 is known to exhibit a complex structure, with five main velocity components as illustrated in Fig. \ref{fig:lines}. At low galactic latitude ($l, b = 150.3772^{\circ}, -1.6037^{\circ}$), the path through the Galactic disk is long, with most of the extinction located at a distance of about 1~kpc and a total reddening of E(g-r) $\sim1.2$\,mag ($A_V \sim 4.5$\,mag ) according to the 3D dust distribution established from the Panstarrs and GAIA data \citep{Green2019}. The five features detected in HCO$^+$ absorption have comparable integrated opacities, hence comparable
HCO$^+$ and H$_2$ column densities, $N(\HCOp) \sim 1.2 \times 10^{12}$ cm$^{-2}$ and $N(\Ht) \sim 4 \times 10^{20}$ cm$^{-2}$, corresponding to $A_V \sim 0.8$\, mag each  \citep{ll96-1} for a total column density integrated in these five velocity components of $N(\Ht) = 2.1 \times 10^{21}$ cm$^{-2}$.  By
comparing HI absorption and emission spectra toward this source, \citet{ll-2000} have derived the atomic hydrogen column density in the velocity range between $-20$ and $0$ kms$^{-1}$, N(HI)=$4.56 \times 10^{21}$cm$^{-2}$. The mean fraction of hydrogen in molecular form along the line of sight is therefore about 0.5.
Taken together, these five absorption features therefore contribute most of the extinction toward NRAO150.

\citet{pety:08} extensively studied the environment of this sightline combining observations with the IRAM 30m single dish telescope and PdBI interferometer. The CO column densities of each velocity component remain moderate, as expected for diffuse gas where
the free gas phase carbon is either C$^+$  for the broader and fainter velocity components, or on the verge of recombining to C and CO for the strong, narrow velocity components, at $-17$ and $-10$ \kms, where the CO column density gets larger than 10$^{16}$cm$^{-2}$. 
 The CO emission profiles show strong
spatial variations with the achieved angular resolution. \citet{pety:08} show that overall spatial variations in the strongly CO emitting gas have Poisson statistics with rms fluctuations about equal to the mean emission level in the line wings and much of the line cores.  The brightest emission peaks reach the brightness temperature of 13 K and the ratio between the CO\Jtwo \,and CO\Jone \, is about 0.7
It is remarkable that the CO emission line profile converges toward the CO absorption profile only when observed at the highest angular resolution of 6''. At lower angular resolution, the CO profiles are broader and more similar to those of HCO$^+$. The variations of the CO emission are explained by the presence of density fluctuations along the line of sight, that affect the CO abundance and excitation, hence introducing variations in the line profile.  As the fraction of hydrogen in molecular form is mostly sensitive to the total gas column, the variations of the molecular hydrogen column density
are expected to be much less than those of an individual molecule \citep{Shull:21,Liszt2023}.

\citet{Araya2014} confirmed the variation of the $-10.4$\kms absorption feature with time in the ground state o-H$_2$CO($1_{1,0}-1_{1,1}$) line at 4.8~GHz.
They interpreted this as a slow increase in the path length with time when the sightline crosses a denser filament  close to its edge. 
They determined a filament radius of $2\times 10^4$ au for a density of $n(\Ht) = 10^4$~cm$^{-3}$, consistent with the CO emission maps \citep{pety:08}. The detection of absorption lines from excited levels of CS,
c-C$_3$H$_2$ (cyclopropenylidene), and H$_2$CO allows to reconsider the density determinations of the different velocity components.

\section{Modeling the molecule excitation}
\Tabcol
We used the RADEX code \citep{vdtak} to study the excitation conditions. The references for the collision rate coefficients are given in Table \ref{tab:cols}. We included collisions with ortho and para H$_2$, electrons and Helium to represent the diffuse and translucent gas where atomic neutral and ionized carbon 
are the main reservoirs of gas phase carbon, up to
the regions where CO dominates the carbon budget. 
All calculations were performed using the default value for the escape probability formulation, a uniform sphere. Changing for a different geometry, for instance a plane-parallel slab,  leads to similar conclusions but introduces changes of the derived densities by $\sim 30\%$. Changing the hypotheses on the electron fraction, H$_2$ ortho-to-para ratio or kinetic temperature also have an impact on the derived densities. We therefore have not attempted to obtain a precise determination of the density, but we chose to bracket the acceptable interval of the molecular hydrogen density, given the restricted set of hypotheses.
 
In the RADEX calculations, we did not include atomic hydrogen as a collision partner as no such data are currently available for the considered molecules. \citet{daniel:2015} have shown in the case of water vapor, that the rate coefficients with atomic hydrogen can be larger or smaller
than those for H$_2$, but they are usually similar in magnitude, at least for the 
dominant transitions with the highest values of the cross sections. The cross sections with H cannot be easily
extrapolated from those with either He or H$_2$ as a collision partner in the case of H$_2$O. For the considered diffuse and translucent clouds, especially for the
narrow velocity components that have the highest abundance of molecular species including CO, molecular hydrogen is expected to dominate the
hydrogen budget. To complete the picture of molecule excitation in diffuse and translucent gas, cross sections with H would be interesting, especially for 
 regions where atomic hydrogen is  dominant.
We assume the electrons are thermalized at the kinetic temperature of the gas as the time scales for thermalization by interactions with electrons or neutrals are shorter than one year in the considered conditions \citep{tielens}.

We considered five cases for the kinetic temperature, 10, 15, 20, 30 and 60~K to bracket the 
possible range of conditions along this sight line which is expected to be similar to the C$_2$ clouds investigated by \citet{Fan:2024}, and varied the electron fraction $x(e)$ between $2\times 10^{-6}$ and $2\times 10^{-4}$ to probe the transition between diffuse and translucent gas. Ten values of the molecular hydrogen density have been considered, 200, 300, 500, 1000, 1500, 2000, 2500, 3000, 5000 and 10000 cm$^{-3}$.

\subsection{CS excitation}
The calculations were performed using the molecular parameters derived from the observed line profiles, 
the peak opacity, velocity dispersion, and column density. For this last parameter, we used the excitation temperature listed in table \ref{tab:fits}, to take into account
the slight deviation from the excitation by the Cosmic Microwave Background (CMB)  for the narrow velocity components at $-17$ and $-10$ kms$^{-1}$. 
We  computed the H$_2$ ortho-to-para ratio  at thermal equilibrium for the kinetic temperature of the medium. Figure \ref{fig:cs} presents the calculations of CS excitation for four velocity components. It displays the variations of the peak opacity of the CS(2-1) and CS(3-2) lines as a function of the molecular hydrogen density, for five values of the kinetic temperature and six values of the electron fraction. Each velocity component is displayed in a different box, from top to bottom, 
$-17.2$~\kms 
, $-10.3$~\kms 
, $-8.5$~\kms 
, and $-4$~\kms . 
The measured opacities for these four velocity components are shown with dashed and dotted  areas.  As listed in Tab. \ref{tab:fits} the strongest velocity components at $-17.2$~\kms \, and $-10.3$~\kms \, show narrow and deep absorption features in the two CS transitions, $\tau \sim 1$ \& FWHM $\sim 0.4$~\kms. The detection of a significant opacity of the   $J=3-2$ transition implies a relatively high H$_2$ density of n(H$_2$) $\sim 2500$~cm$^{-3}$ for these two velocity components. As indicated by the H$_2$CO excitation described below, these two velocity components have a moderate electron fraction, $x(e) < 10^{-4}$, and the CS excitation is mainly dependent on the gas density and temperature. 
 The faint components at $-8.5$ and $-4$ ~\kms \, exhibit broader profiles  ($\tau \sim 0.1$ \& FWHM $\sim 1.5$~\kms) and a lower level of excitation as compared to the main components. For these two velocity components, the low CS(3-2) opacities are consistent with the typical densities encountered in diffuse and translucent clouds, n(H$_2$) $\leq 500$~cm$^{-3}$ and any electron fraction.
At such densities, the excitation temperature of CS, as that of other high-dipole moment species is very close to the temperature of the CMB.

\subsection{H$_2$CO excitation}
We also performed excitation calculations for ortho and para formaldehyde, using the extended set of physical conditions with five values of the kinetic temperatures,  six values of the electron fraction, 
and five values of the H$_2$ ortho-to-para ratio, $opr$ = $3\times 10^{-5}$,  $3\times 10^{-4}$, $0.03$, $0.3$ and $3$ for the five values of the kinetic temperature.
We choose a value for the o-H$_2$CO column density,
$2.2 \times 10^{12}$~cm$^{-2}$, and line width (0.5 \kms) corresponding to the measured column density in the $-17$\kms\, velocity component. 
  The calculations have been done for different values of the volume density, but we only show the results for  $n(\Ht) = 10^3$~cm$^{-3}$ in Fig. \ref{fig:h2co}, a value intermediate between the density of the broad and the narrow velocity components. The behavior is qualitatively similar for other densities in the range of the densities encountered
 in the  clouds along this line of sight. To illustrate the general behavior of the excitation 
 of the rotational lines of formaldehyde, and provide information for future observations, we have included in the calculations two low energy lines of ortho and 
 para formaldehyde that can be easily observed with ground based millimeter telescopes, the $2_{1,2} - 1_{1,1}$ line of o-\HHCO \, at 140~GHz and the
 ground state $1_{0,1} - 0_{0,0}$ transition of p-\HHCO \, at 72~GHz. The extensive set of calculations is presented in Appendix \ref{sec:appendix}.

 The $1_{1,0}-1_{1,1}$ transition of \HHCO \, at 
4.8~GHz is particularly sensitive to the physical conditions. In the cold envelopes of molecular clouds, it can have an excitation temperature lower than the CMB because of specific propensity rules
for the collision of o-\HHCO \, with H$_2$ molecules \citep{Troscompt:2009}. The effect depends on the ortho-to-para ratio of H$_2$, since collisions with o-H$_2$ are more efficient at 
exciting this transition leading to a smaller range of physical conditions where the transition is cooled when using collisions with ortho and para H$_2$, than when using collisions with p-H$_2$ only. Collisions with electrons can also have an important effect for the diffuse regions
we probe, as noted previously by \citet{Kutner} and by \citet{Turner}. 
These authors used crude approximate analytical formulations for the excitation cross sections with electrons. New cross sections have therefore been computed here using the Born dipole theory which works well for dipolar transitions in molecules with large dipoles 
(the H$_2$CO dipole is 2.3 D). Rate coefficients were then obtained by integrating the cross sections over Maxwell-Boltzmann distributions of the electron velocity for kinetic temperatures in the range 10-300 K. This dataset will be made available through the 
Excitation of Molecules and Atoms for Astrophysics (EMAA) database \footnote{ https://emaa.osug.fr and https://dx.doi.org/10.17178/EMAA}.


 Figure \ref{fig:h2co} presents the results for three transitions of o-H$_2$CO,  the $1_{1,0}-1_{1,1}$ line at 4.8~GHz, together with the two strong transitions which relate to the levels involved in the 4.8~GHz line, the $2_{1,2}-1_{1,1}$ line at 140~GHz (not observed in this work), and the $2_{1,1}-1_{1,0}$ line at 150~GHz which was observed at NOEMA. Over most of the covered set of physical conditions, the excitation temperature of the latter two lines remain close to the CMB, except for the highest values of the density and electron fraction,  and the opacity is almost constant. Fig. \ref{fig:goh2co} shows the variation of the excitation temperature  with the molecular gas density. The excitation temperature and opacity show a small dependence on the kinetic temperature and H$_2$ ortho-to-para ratio with a higher excitation temperature at higher kinetic temperature and $opr$.
 At low values of the electron fraction($< 3 \times 10^{-5}$), a small asymmetry between the 2mm lines appears with slightly higher values for the 140~GHz line opacity and slightly lower values for the 150~GHz line opacity as compared with the same set of parameters and a higher electron fraction. This effect is a consequence of the strong cooling of  the excitation temperature of the 4.8~GHz line which is  predicted to be significantly lower than the CMB,  that is to say overcooled. 
 For the chosen column density, density and line width, the overcooling of the 4.8~GHz line  is present up to an electron fraction of about $6\times 10^{-5}$ depending on the kinetic temperature and H$_2$ $opr$. As illustrated in Fig.\ref{fig:h2co} (top left and top right panels) and Fig.\ref{fig:goh2co}, regions with lower electron densities, lower $opr$  and higher kinetic temperatures seem to be favorable for the  overcooling phenomenon within the range of parameters covered. The opacity of the  $1_{1,0}-1_{1,1}$ transition is strongly affected by its excitation conditions : the region with overcooling is associated with a significant opacity while the opacity decreases when  the excitation temperature increases above the CMB. This strong sensitivity of the 4.8~GHz line excitation temperature to the local physical conditions, for a given value of  the molecular column density may qualitatively explain the higher rate of  variations of the line profile with time toward quasar sight lines, as compared with other species \citep{Araya2014}.  Indeed, a variation of the electron fraction, which could be combined with a variation of the local density, or the kinetic temperature, or of the H$_2$CO column density when the line of sight is crossing a filamentary structure could explain the detected trend in the H$_2$CO profile toward NRAO150 \citep{Araya2014},  and the absence of significant variation in other species such as HCO$^+$ \citep{ll-2000}. The sensitivity of the line profile to the excitation conditions implies that , for a given H$_2$CO column density, regions with a moderate electron fraction will be more easily detected because the line opacity with be higher. Such regions will appear more prominently in the 4.8~GHz line profile than regions with a high electron fraction. By contrast, the 2~mm lines are mostly sensitive to the molecular column density  and do not determine the physical conditions. These lines are therefore more appropriate for studying the formaldehyde column density and abundance.
 
 According to the excitation calculations shown in Fig. \ref{fig:ph2co}, the 145 GHz p-H$_2$CO line optical depth should be less affected by electrons than the 150 GHz o-H$_2$CO line (compare Fig. \ref{fig:h2co} and Fig. \ref{fig:ph2co}). We list in Table \ref{tab:ratio} the ratios of integrated opacities of the discussed H$_2$CO lines with respect to the
 measurement of the strong component at $-10$~\kms.   The high level of agreement between the line ratios in the two mm-wave lines
 indicates that the $-10$~\ kms \, line optical depth has not been strongly affected by electrons. The ~10\% difference between the mm-lines at 
$-17$~\kms \, indicates a higher electron density at $-17$~\kms \, than at $-10$~\kms \, but less than at $-14$, $-8$ or $-4$~\kms. 
The intrinsic column density ratio  N(v=-17)/N(v=-10) might then be given by the ratio of the p-H$_2$CO column density and be about 1/1.72
 (see Table \ref{tab:ratio}).
 Using the excitation calculations, we conclude that the electron fraction in the two narrow components is about  $x(e) \sim 1 - 3\times 10^{-5}$, a value consistent with the CS calculations, noting that the electron fraction is higher in the $-17$~\kms \, component. 
 The electron fraction of the broad components  must be significantly  higher since these components are much less prominent at 4.8~GHz than at 150~GHz. In particular the comparison of the spectra traced in blue on the left and right panels of Fig.~\ref{fig:lines} shows that the velocity components at $-13.8$ and $-4.2$ kms$^{-1}$ are not detected and that at $-8.4$ is barely seen.

In the range of densities and electron fraction derived from the CS and o-H$_2$CO excitation analysis, 
the excitation temperature of the 2~mm lines remain close to the CMB and does not exceed 3~K. Similar calculations for p-H$_2$CO lead to the same conclusion.
Hence the derivation of the formaldehyde column densities from the integrated opacities is straightforward using the simple assumption that the molecules are thermalized with the CMB.

\Tabratio

\subsection{c-C$_3$H$_2$ excitation}
Excitation calculations are reported in appendix \ref{sec:appendix} and illustrated in Fig. \ref{fig:c3h2}. Given the high dipole moment of this molecule, 3.27 Debye \citep{Lovas:1992} and the good correspondance of the line profiles with those of CCH and HCO$^+$ with no narrow and deep features, it is expected that the excitation temperature of its millimeter rotational lines is close to the CMB. Excitation calculations with RADEX confirm the hypothesis.

\section{Molecular column densities}
\label{sec:col}

We used the information on the excitation conditions to derive the column densities for the five  main  velocity components. They are listed in the last column of Tab.\ref{tab:fits} with the assumed excitation temperature. The new data confirm the previous results \citep{ll95} but  provide better determinations due to the increased sensitivity of NOEMA as compared with 
the Plateau de Bure Interferometer. The new data also provide contraints on the molecular gas density and electron fraction for the velocity components.
Tab. \ref{tab:abund} summarizes the abundances relative to H$_2$ of the studied molecules, CS, \HHCO \, and c-C$_3$H$_2$, and the ortho-to-para ratio for the latter two species for
each velocity component separately.
For all velocity components, we use the HCO$^+$  integrated opacity to determine the H$_2$ column density assuming the fixed abundance relative to H$_2$, [HCO$^+$] $=  3 \times 10^{-9}$ \citep{Gerin:2019}. Abundances of the rare isotopologues H$_2$$^{13}$CO and C$^{34}$S are also listed for the velocity components where these species are detected. 

The CS abundances are four times higher in the narrow, higher density velocity components than in the more diffuse components, reaching $\sim 10^{-8}$, a similar or even higher abundance than in nearby molecular clouds, $ 10^{-9.5}$ to 10$^{-7.5}$ \citep{Tafalla:2021,Rodriguez:2021}. In those velocity components C$^{34}$S is detected and the [CS]/[C$^{34}$S] abundance ratio is $22 \pm 5$, consistent with the sulfur isotopic ratios in the solar neighborhood, 22 \citep{wilson,yan:2023}. The large increase of the CS abundance indicates a more efficient formation mechanism at the higher densities encountered in the narrow velocity components. Even in the low density components the CS abundance remains above 10$^{-9}$, a significant level for such highly illuminated and low density conditions.

For formaldehyde, the ortho-to-para ratio  is consistent with the statistical value of three in all velocity components. 
Formaldehyde abundances are also higher in the velocity components associated with higher gas densities but the  variation is not as pronounced as that of CS, from $\sim 2.5 \times 10^{-9}$ in the low density components up to $\sim 6.7 \times 10^{-9}$ in the two higher density components, which represents an increase of a factor of 2.7. The same range of increase of the abundance has been seen for CO \citep{ll98} although the strongest component in \HHCO\, and in CO are different.
This increase of the abundance implies that the  chemistry   is somewhat more efficient at producing \HHCO\,  in the higher density velocity components than for the low density components.
o-H$_2$$^{13}$CO is detected in the strongest narrow velocity component near $-10$ \kms. The corresponding $^{12}$C/$^{13}$C ratio is $61 \pm 12$ indicating that o-\HHCO \, is not heavily fractionated, contrasting with  CO along the same sightline where $^{12}$CO/$^{13}$CO$=25$ \citep{ll98} for this particular velocity component. The full range 
of the $^{12}$CO/$^{13}$CO abundance ratio extends from 15 to 33 for the five velocity components. CO and \HHCO \, therefore seem to have different chemical pathways.

The c-C$_3$H$_2$ abundance relative to H$_2$ does not vary much across  the five velocity components with a possible trend of higher values for the more diffuse, lower density components with broader profiles. The mean abundance of ortho c-C$_3$H$_2$ is  $1.8 \pm  1 \times 10^{-9}$. As for \HHCO, the ortho-to-para ratio is generally consistent with the statistical value of three.
The derived abundance toward NRAO150 is comparable with the abundance detected in the Horsehead photodissociation region, $1.3 \times 10^{-9}$ \citep{pety:2012} and the mean abundance of o-C$_3$H$_2$ in diffuse clouds, $1.4 \pm 0.7 \times 10^{-9}$ \citep{ll00}.
By comparison, the abundance of ortho c-C$_3$H$_2$ is $\sim 4.5 \times 10^{-9}$ in the TMC-1 dark cloud \citep{cernicharo:2021} and $\sim 5 \times 10^{-9}$  in L483 \citep{Agundez:2019}.   There are few measurements of the ortho-to-para ratio but for TMC-1, it seems consistent with three \citep{cernicharo:2021}. The abundance of
c-C$_3$H$_2$ therefore increases by a factor of $\sim 3$ in dark and dense clouds  as compared with UV-illuminated diffuse and translucent clouds. While the increase in CS and H$_2$CO abundance is already seen in the two moderately dense components along the line of sight to NRAO~150, these components do not stand out 
in c-C$_3$H$_2$. The chemical changes seem therefore to require a more pronounced modification of the environment, such as higher densities and extinctions than those encountered along this particular line of sight.

\Tabab

\section{Comparison of the H$_2$CO abundance and excitation in other regions }
\label{sec:comp}
Large-scale maps of the formaldehyde 4.8~GHz  have been obtained toward molecular clouds and complexes. Extended absorption is detected in this line, which remains optically thin.
The high latitude molecular cloud MBM40 has been studied by  \citet{Monaci:2023} in HI, $^{12}$CO(1-0), $^{13}$CO(1-0),  and 4.8~GHz o-H$_2$CO among other lines. 
The o-H$_2$CO line is detected in absorption and its profile is similar to that of $^{13}$CO (1-0) and narrower than that of $^{12}$CO and HCO$^+$(1-0). 
\citet{Gong:2023} present large-scale maps of the o-H$_2$CO line at 4.8~GHz toward the Cygnus  star forming molecular complex. The line is detected in absorption over very large scales, indicating an excitation temperature lower than the CMB. As for MBM40, the $1_{1,0}-1_{0,1}$ integrated opacity is well correlated with the $^{13}$CO(1-0) emission. 
This behavior can be explained by our excitation calculations since the opacity of the o-H$_2$CO line is stronger for the denser and lower electron fraction regions which are associated with narrower line profiles. Absorption at 4.8~GHz preferentially selects the regions along the line of sight with a low electron fraction and a relatively low \Ht\ $opr$.  
\citet{Gong:2023} performed RADEX excitation calculations to evaluate the behavior of the 4.8~GHz line and to derive the o-H$_2$CO column densities,  using collision cross sections with ortho and para \Ht, but a single value of the ortho-to-para ratio and a limited range of kinetic temperature. As neither the influence of the \Ht\ $opr$ nor the role of electron excitation was included, the derived column densities may have been under-estimated, leading to a significantly lower abundance relative to H$_2$  than that we derived, $7\times 10^{-10}$ as compared with $2 - 8 \times 10^{-9}$.

\section{Chemistry}
\label{sec:chem}
The analysis of the excitation conditions shows that the different velocity components have somewhat different densities, ranging from a few hundred cm$^{-3}$ for the
broad velocity components at $-13$, $-8$ and $-4$ kms$^{-1}$, up to $\sim 2500$ cm$^{-3}$ for the narrow components at $-17$ kms$^{-1}$ and $-10$ kms$^{-1}$. 
  \HHCO\, is present in all  velocity components, with a higher abundance relative to H$_2$ in the higher density velocity components. At the moderate extinction along this sightline (Av = 4.5 magnitude in total and 0.8 mag for each velocity component),  grains are expected to stay bare with little condensation of water ice, especially in the low density, diffuse components.  
  The most likely chemical formation mechanism of formaldehyde is therefore the gas-phase
reaction : $\emr{O}+\CHHH \longrightarrow \emr{H}+\HHCO$ with a rate of about $\sim 1.1 \times 10^{-10}$ cm$^3$s$^{-1}$ at 300~K in the KIDA data base\footnote{https://kida.astrochem-tools.org} \citep{Wakelam:2012}. This reaction has been 
evaluated in the framework of combustion and atmospheric chemistry \citep{Baulch:05,Atkinson:06}. \citet{Ramal:21} performed theoretical calculations on the reactions forming formaldehyde in interstellar clouds. They showed that the reaction between O and CH$_3$ can be effective in diffuse cloud conditions. Other reactions may also contribute to the production of gas phase formaldehyde, involving O, O$_2$ or OH, and hydrocarbons as reactants, as well as dissociative recombination reactions of complex organic ions with electrons.
Formaldehyde is destroyed by photons, cosmic rays and reactions with atoms and ions, notably C and C$^+$. 
Because most of the gas phase oxygen is expected to be neutral and atomic in diffuse clouds, this chemistry may
constrain the abundance of the methyl radical \CHHH{} if the reaction rate is known with a reasonable precision, of a  factor of two or better.
The observed increase by a factor 2.5  of the H$_2$CO abundance with the density is qualitatively consistent with a gas phase production route, since the formation
rate of formaldehyde by reactive collisions increases more rapidly with the molecular hydrogen density than its destruction rate by photons or cosmic rays. Hence the formaldehyde abundance may be higher in denser gas, for a given molecular hydrogen column density and extinction. Quantitative models are needed to fully test the formation and destruction processes of formaldehyde and how its relative abundance scales with the molecular hydrogen density and electron fraction.


\section{Conclusions}
\label{sec:conc}
We presented new observations of the sightline toward NRAO150 with the NOEMA interferometer. The detected absorption lines from \HHCO, CS, and c-C$_3$H$_2$
were combined with existing data at lower frequencies,  and used to constrain the physical conditions along this sightline and the  chemical mechanisms forming \HHCO. Cross sections for the collisional excitation of o-H$_2$CO and p-H$_2$CO with electrons have been computed allowing a detailed analysis of their excitation
  in diffuse and translucent cloud conditions, with a focus on the behavior of the $1_{1,0}-1_{1,1}$ transition of \HHCO\, at 4.8 GHz. 
  These cross sections are available in the EMAA data base. Cross sections for the collisional excitation of CS and H$_2$CO with atomic hydrogen are missing and
  could be of interest for the study of these molecules in the most diffuse regions. 

\begin{itemize}
\item The excitation of the $1_{1,0}-1_{1,1}$ transition of \HHCO\, at 4.8 GHz is sensitive to the electron fraction, in addition to the kinetic temperature and ortho-to-para ratio of H$_2$. 
This line therefore does not provide by itself a measurement of the \HHCO\ column density. Combined with observations of  the higher frequency rotational lines of \HHCO\ at 3mm or 2mm, this line can be used as a diagnostic of the electron fraction and H$_2$ ortho-to-para ratio. 
\item The narrow components at $-17$ \rev{kms$^{-1}$} and $-10$ kms$^{-1}$ correspond to moderately dense gas, n(H$_2$) $\sim 2500$ cm$^{-3}$, at the transition from C$^+$ to C and  CO, with an electron fraction of $\sim 3 \times 10^{-5}$ where the neutral carbon atom is the main bearer of carbon.
\item The broad components at  $-13$, $-8$, and $-4$ kms$^{-1}$ correspond to low density gas (n(H$_2$) $< 500$ cm$^{-3}$), with a high electron fraction ($\sim  10^{-4}$), 
where C$^+$ is the dominant reservoir of carbon. 
\item The CS abundance increases by a factor of four in the narrow velocity components and reaches $10^{-8}$ .
\item The abundance of \HHCO \, increases by a factor of 2.5, from $2.5 \times 10^{-9}$ to $6.5 \times 10^{-9}$, from the low density to the moderate density components. 
\item The abundance of c-C$_3$H$_2$ is equivalent in all velocity component [c-C$_3$H$_2$] = $1.8 \times 10^{-9}$.
\item We derive ortho-to-para ratios of $3$, the statistical value for all components detected in \HHCO \, and in c-C$_3$H$_2$. 
\item We derived the isotopic ratios [\HHCO]/[H$_2^{13}$CO] = $61\pm 12$ and [CS]/[C$^{34}$S] =$24 \pm 6$ in the narrow velocity components.
\end{itemize}

\begin{acknowledgements}
This work is based on observations carried out under project number W21AB with the IRAM NOEMA Interferometer. IRAM is supported by INSU/CNRS (France), MPG (Germany) and IGN (Spain). We thank the IRAM staff, 
 for help in the preparation and reduction of the observations and for the high  quality of the NOEMA observations.  
This research has made use of spectroscopic and collisional data from the EMAA database (https://emaa.osug.fr and https://dx.doi.org/10.17178/EMAA). EMAA is supported by the Observatoire des Sciences de l’Univers de Grenoble (OSUG). 
This work was supported by the program "Physique et Chimie du Milieu Interstellaire (PCMI)" funded by Centre National de la Recherche Scientifique (CNRS) and Centre National d'Etudes Spatiales (CNES). 
The National Radio Astronomy Observatory is operated by Associated Universities, Inc. under a cooperative agreement with the National Science Foundation.

\end{acknowledgements}

\bibliographystyle{aa} 
\bibliography{h2co} 

\appendix
\section{RADEX calculations}
\label{sec:appendix}
\FigGRP

\FigGRO

In this appendix we present more extensive calculations of o-H$_2$CO and p-H$_2$CO excitation using the new set of cross sections including electron excitation. 

Fig. \ref{fig:goh2co} and Fig.\ref{fig:gph2co} present the variation of the excitation temperature and opacity of the low energy transitions of o-H$_2$CO and p-H$_2$CO.
The grid covers the same parameter space as the CS calculations, five values for the kinetic temperature, 10, 15, 20, 30 and 60~K, ten values for the molecular hydrogen density, 200, 300, 500, 1000, 1500, 2000, 2500, 3000, 5000, and 10000 cm$^{-3}$, and six values for the electron fraction, $2\times 10^{-6}$,  $10^{-5}$, $3\times 10^{-5}$, $6\times 10^{-5}$, $10^{-4}$, and $2\times 10^{-4}$. The chosen range encompasses the low electron fraction values where electron excitation is negligible and the high electron fraction values where electron excitation plays a significant role. The chosen column densities are identical as those discussed in Fig.\ref{fig:h2co} and Fig.\ref{fig:ph2co}, namely N(o-H$_2$CO)$=2.2\times 10^{12}$~cm$^{-2}$ and  N(p-H$_2$CO)$=7.3\times 10^{11}$~cm$^{-2}$, and the FWHM is set at $\Delta V = 0.5$\kms.

The behavior of the $1_{1,0}-1_{1,1}$ line is very different from that of the other low energy transitions of either o-H$_2$CO or p-H$_2$CO, which involve levels with a larger difference of energies and frequencies larger than 70~GHz. For those transitions, the excitation temperature is always higher than the CMB, and a higher excitation is mainly governed by the higher values of the density and/or the kinetic temperature. The electrons provide an additional source of excitation for high values of the electron fraction, typically above an electron fraction of $\sim §\times 10^{-5}$, corresponding to the transition region between the three main carbon reservoirs, C$^+$, C and CO. The excitation temperature of the  $1_{1,0}-1_{1,1}$ line takes values lower than the CMB when the electron fraction is low to moderate, and stays below  $\sim 5\times 10^{-5}$ for the range of parameters covered here.  The excitation temperatures lower than the CMB are also favored by higher values of the kinetic temperature and low values of the \Ht \, ortho-to-para ratio as shown in Fig \ref{fig:h2co}. Variations of the excitation temperature are associated to variations of the corresponding line opacity for a given column density.  Even when the line is detected in absorption, the range of excitation temperature and corresponding opacities is significant,  and may reach a factor of four. Therefore, column density determinations based on the $1_{1,0}-1_{1,1}$ line at 4.8~GHz should be based on an accurate
knowledge of the physical conditions along the line of sight, including the electron fraction and kinetic temperature.

\FigPH

Figure \ref{fig:ph2co} presents the variation of the p-H$_2$CO lines with the electron fraction and H$_2$ ortho-to-para ratio in the same way as Fig. \ref{fig:h2co} for o-H$_2$CO. The molecular hydrogen density is fixed at n(\Ht)=10$^3$~cm$^{-3}$, the p-H$_2$CO column density is  N(p-H$_2$CO)$=7.3\times 10^{11}$~cm$^{-2}$, and the line width is  $\Delta V = 0.5$\kms. The excitation of the p-H$_2$CO line is similar to that of the 2~mm lines of o-H$_2$CO and exhibits a simple variation with the electron fraction, \Ht \, density, $opr$ and gas kinetic temperature.  For diffuse and translucent clouds with moderate densities, the excitation temperature of the two lines
remains within 0.5~K from the CMB for \Ht \, densities  lower than $\sim 2000$~cm$^{-3}$. The derivation of p-H$_2$CO column densities from integrated opacities is therefore straightforward even when few informations on the physical conditions along the line of sight are accessible.
 
\FigC

Figure \ref{fig:c3h2} shows RADEX calculation for ortho c-C$_3$H$_2$, using a total column density of 10$^{12}$~cm$^{-2}$ and a line FWHM of 0.6 kms$^{-1}$. These calculations use collisional cross sections with para H$_2$ and He only \citep{Khalifa:2019}. The effect of collisional excitation  is more pronounced for the lowest frequency transition, 
the $1_{1,0}-1_{0,1}$ line at 18~GHz, than for the other lines at higher frequencies. For these lines, the CMB is the main source of excitation up to densities larger than about 2000~cm$^{-3}$, as expected for such a high dipole moment species.

\end{document}